\documentclass[preprint,12pt]{elsarticle}

\usepackage{amssymb}
\usepackage{amsmath}

\usepackage{booktabs,siunitx}
\usepackage{multirow}
\usepackage{url}

\usepackage{tikz}

\def\H{{\mathbf H}}

\def\W{{\mathbf W}}
\def\A{{\mathbf A}}

\def\M{{\mathbf M}}

\def\bR{{\mathbb R}}

\newcommand{\mb}[1]{\mathbf{#1}}

\newcommand{\cat}[2]{\underset{#1} {\overset{#2} {\Large \mathbin\Vert}}}

\def\mha{{f_{\mathrm{MA}}}}
\def\att{{f_{\mathrm{AT}}}}
\def\LN{{\mathrm{LN}}}

\journal{Computer Speech \& Language}

\begin{document}

\begin{frontmatter}



\title{DiCoW: Diarization-Conditioned Whisper for Target Speaker Automatic Speech Recognition}

\author[label1]{Alexander Polok\corref{cor1}%
\fnref{contrib}}
\ead{ipoloka@fit.vut.cz}
\author[label1,label2]{Dominik Klement\fnref{contrib}}
\author[label1]{Martin Kocour\fnref{contrib}}
\author[label1]{Jiangyu Han}
\author[label1]{Federico Landini}
\author[label1]{Bolaji Yusuf}
\author[label2,label3]{Matthew Wiesner}
\author[label2,label3]{Sanjeev Khudanpur}
\author[label1]{Jan Černocký}
\author[label1]{Lukáš Burget}

\fntext[contrib]{These authors contributed equally.}
\cortext[cor1]{Corresponding author}

\affiliation[label1]{organization={Speech@FIT, Brno University of Technology},
            country={Czechia}}

\affiliation[label2]{organization={CLSP,Johns Hopkins University},
            country={USA}}
\affiliation[label3]{organization={HLTCOE, Johns Hopkins University},
            country={USA}}


\begin{abstract}

Speaker-attributed automatic speech recognition (ASR) in multi-speaker environments remains a significant challenge, particularly when systems conditioned on speaker embeddings fail to generalize to unseen speakers. In this work, we propose Diarization-Conditioned Whisper (DiCoW), a novel approach to target-speaker ASR that leverages speaker diarization outputs as conditioning information. DiCoW extends the pre-trained Whisper model by integrating diarization labels directly, eliminating reliance on speaker embeddings and reducing the need for extensive speaker-specific training data.
Our method introduces frame-level diarization-dependent transformations (FDDT) and query-key biasing (QKb) techniques to refine the model's focus on target speakers while effectively handling overlapping speech. By leveraging diarization outputs as conditioning signals, DiCoW simplifies the workflow for multi-speaker ASR, improves generalization to unseen speakers and enables more reliable transcription in real-world multi-speaker recordings.
Additionally, we explore the integration of a connectionist temporal classification (CTC) head to Whisper and demonstrate its ability to improve transcription efficiency through hybrid decoding. Notably, we show that our approach is not limited to Whisper; it also provides similar benefits when applied to the Branchformer model.
We validate DiCoW on real-world datasets, including AMI and NOTSOFAR-1 from CHiME-8 challenge, as well as synthetic benchmarks such as Libri2Mix and LibriCSS, enabling direct comparisons with previous methods. Results demonstrate that DiCoW enhances the model's target-speaker ASR capabilities while maintaining Whisper's accuracy and robustness on single-speaker data.
\end{abstract}


\end{frontmatter}
\section{Introduction}
\label{sec:intro}
    
The rapid development of deep learning techniques and vast increases in available training data and computing resources have made low-error-rate ASR systems on single speaker recordings viable at reasonable latencies ~\cite{li2022recent}.

Consequently, the research community has focused~\cite{watanabe2020chime,Yu2021M2MetTI,cornell2023chime,cornell2024chime} on the more challenging task of multi-speaker ASR, whose goal is the accurate tran\-scrip\-tion of multiple speakers in a recording, often including speaker-attributed ASR, whose goal is also to assign a speaker label to each spoken word.

In this work, we focus on speaker-attributed ASR by adapting a single-speaker ASR system using speaker diarization information to produce transcripts for all speakers.

Speaker-attributed ASR-systems that combine diarization and speech recognition mainly operate in one of three ways: (a) running ASR and diarization independently, capturing the respective word and timing information from each, and using this timing information to assign words to speakers throughout a long conversation~\cite{bhandari2024reverb}; (b) using a fully cascaded pipeline where various orderings of diarization, speaker extraction or source separation are combined with ASR~\cite{yoshioka_19_meeting_transcription,desh_21_css_multitalker}; and (c) using target speaker ASR (TS-ASR), where, in lieu of any source extraction or separation, the original audio is directly input along with some speaker conditioning, so that the system transcribes the speech belonging to the conditioned speaker~\cite{Kanda2019_spkloss}. 

The conventional approach to TS-ASR is to extract speaker embeddings corresponding to the target speaker and to have these embeddings as an auxiliary input to the ASR system~\cite{karafiat2011ivector,Zili23_adapting}. 
Although the use of speaker embeddings from a pre-trained speaker embedding extractor~\cite{dehak2010front,snyder2018x,wang2023wespeaker} can give clues to the ASR about what information in its input to utilize and what information to ignore, it implicitly requires learning to map speaker embeddings to ASR speech embeddings. 
Learning a robust mapping ge\-ne\-ra\-lizing to new speaker sets at test time, especially when the speaker em\-be\-ddings are trained independently, requires having a large number of speakers in the multi-speaker ASR training set. While this is manageable for si\-mu\-lated multi-speaker data (for which a large number of speakers could be obtained), data scarcity is arguably the most significant challenge for ``in-the-wild" multi-speaker ASR; it is therefore, imperative to develop systems that can be efficiently trained on the order of tens of hours of real conversational data.

In this paper, we propose Diarization-Conditioned Whisper (DiCoW), a semi-end-to-end approach to speaker-attributed ASR where we condition Whisper~\cite{radford2023robust} on diarization outputs, unlike prior approaches relying on speaker embeddings or specific modules to model speaker information. We use Whisper as the base ASR system in order to take advantage of its large-scale pretraining, multi-domain robustness, and long-form capabilities. However, we also present results with another ``generic'' single-speaker system, showcasing that our proposed method can attain strong results in combination with different ASR models.
By directly using diarization labels, there is no need for the model to learn to map speaker and ASR embedding subspaces. This is accomplished by means of speaker conditioning in the form of time-speaker activity probabilities.

To this end, we propose a pair of methods for incorporating speaker ac\-ti\-vi\-ty information into Whisper.

The first of our methods, termed Query-Key Biasing, produces a mask for each target speaker from the diarization outputs.
This mask is then used to modify the attention scores: the scores for frames that do not correspond to the desired speaker are attenuated, while those belonging to the target speaker are kept intact.
This allows Whisper to focus on the target speaker's ASR in the presence of large regions of silence and non-target speaker speech.
We note that the attention scores are modified with trainable parameters so that the attenuation of non-target-speaker positions is only enforced at the beginning of training, and the model is still afforded the flexibility to learn how much non-target information to keep.

In the second method, named Frame-Level Diarization
Dependent Transformations (FDDT), the model is provided with more fine-grained access to the diarization output.
Specifically, for each target speaker, an external diarizer categorizes speech frames into silence, target speaker, non-target speaker and overlapped speech.
For each of the four categories and each Whisper encoder layer, a trainable affine transformation is introduced to transform the input frames belonging to the given category before they are fed into the next encoder layer.
Thus, the model is equipped to learn how to handle the different regions of speech.

We experimentally validate our methods by fine-tuning Whisper on various datasets: NOTSOFAR-1~\cite{vinnikov24_interspeech}, AMI~\cite{Mccowan2005_ami}, and Libri2Mix~\cite{Cosentino2020LibriMixAO} using ground-truth speaker segmentation information. At inference time, we utilize speaker diarization labels generated by an end-to-end speaker diarization system with vector clustering~\cite{kinoshita2021integrating, bredin2023pyannote}. We also evaluate our system on LibriCSS~\cite{chen2020continuous} without fine-tuning it on this dataset.

Our experiments on both real and synthetic datasets show that, without considerably degrading its single-speaker recognition performance, our proposed methods imbue Whisper with strong speaker-attributed ASR capabilities across datasets, even when automatic diarization is used for conditioning.

The rest of the paper is organized as follows:
\begin{itemize}
    \item Section~\ref{sec:related} provides coverage of closely related works.
    \item Section~\ref{sec:longformwhisper} gives a background of Whisper and our modifications to reduce its hallucination tendencies.
    \item Section~\ref{sec:dcwv2} describes the methods we propose to make a diarization-conditioned Whisper capable of target-speaker ASR.
    \item Section~\ref{sec:experiments_setup} describes the setup of our experiments, including datasets, metrics, and training details.
    \item Section~\ref{sec:experiments} shows the results of our experiments.
    \item Section~\ref{sec:conclusions} concludes the paper with a summary of our findings.
\end{itemize}

\section{Related Works}
\label{sec:related}
\textit{Diarization-based ASR:} The integration of ASR and diarization has been explored in various studies using different techniques. These include adding speaker role tokens during ASR decoding~\cite{shafey19_interspeech}, clustering speaker embeddings, and mapping them to ASR tokens~\cite{kanda2022transcribe}---which requires deriving word timings for the tokens after decoding---and jointly producing ASR tokens, speaker tags, and timings~\cite{cornell2024one}. While training a single model from scratch to perform both tasks at once can exploit the synergies between ``what is said" and ``who said it", large training corpora are necessary, which are either only available to large companies or of synthetic nature in academic settings. In contrast, we focus in this work on leveraging existing pre-trained models to reduce the training burden.

\textit{Multi-speaker extensions of Whisper:} There have been recent works extending Whisper for multi-speaker ASR. In~\cite{ma2024extending}, the model is prompt-tuned with a target speaker embedding so that the model recognizes only the speech of that speaker. Alternatively, in~\cite{meng24c_interspeech}, Whisper is activated using speech from the target speaker instead of an embedding so that smaller modifications to the original architecture are required compared to~\cite{ma2024extending}. In~\cite{guo2024sq}, a ``speaker-querying'' module is added to produce speaker prompts that are used as input in the decoder. While similar in motivation, our work differs from~\cite{ma2024extending,meng24c_interspeech,guo2024sq} in that instead of using speaker embeddings or enrollment speech, we directly utilize speaker activities. This simplifies the interaction with external mo\-dules and elides any need for selecting and processing enrollment speech.

\section{Long-form Modeling with Whisper} 
\label{sec:longformwhisper}
OpenAI's Whisper \cite{radford2023robust} is an attention-based encoder-decoder model for automatic speech recognition and speech translation. The widely used model is trained on an order of magnitude more data than other open-source models, which was found to be the key to achieving state-of-the-art performance on a wide range of ASR benchmarks and popularized a number of useful features for ASR.

First, several speech processing tasks need to be performed ``in the wild,'' and it might be relevant to perform them jointly, so Whisper is designed to be prompted with token-based control sequences in order to perform ASR and additionally return time voice activity detection (VAD) or language identification (LID) decisions, among other complementary information.

Whisper incorporates previous text conditioning, where prior transcriptions are fed as context to the decoder. This feature enables effective processing of long-form audio, such as meetings, lectures, or podcasts, by maintaining context across extended recordings.

Long-form audio processing is particularly relevant for multi-speaker ASR, which requires handling continuous dialogues rather than isolated utterances. However, extending Whisper to support multi-speaker scenarios introduces new challenges, such as managing overlapping speech and multi-speaker outputs. Leveraging Whisper as a foundation benefits from its extensive pre-training on large-scale data, reducing the need for additional task-specific training data.

In this work, we propose several extensions to Whisper that address these challenges and enable its application to multi-speaker ASR. The Whisper model is described in the following section, and our proposed methods are detailed in Section~\ref{sec:dcwv2}.

\subsection{Whisper}
\label{sec:whisper_model}
Whisper is available in variants ranging from 38M to 1.54B parameters and has been trained on up to 5 million hours of weakly (pseudo) labeled data. It employs an encoder-decoder Transformer~\cite{vaswani2017attention} architecture, processing the log-Mel spectrogram as input \( \mathbf{X} \in \mathbb{R}^{F \times T} \), where \( T = 3000 \) corresponds to 30 seconds of audio. Shorter audio segments are padded with zero signal, while longer ones are processed sequentially. The number of Mel frequency bins, \( F \), is 80 in earlier versions and 128 in later versions.  

The spectrogram is passed through two 1-dimensional convolutional layers that increase the feature dimension to \( d_m \) and downsample the sequence by a factor of two. The encoder layers transform these features into hidden representations \( \mathbf{H} \in \mathbb{R}^{d_m \times T/2} \), which the decoder uses autoregressively to generate text tokens \( \hat{y} \), conditioned on task-specific special tokens \( g \). The process is formally defined as:  
\begin{equation}
    \mathbf{H} = \text{AudioEncoder}_{\phi_e}(\mathbf{X}), \quad \hat{y}_t = \text{TextDecoder}_{\phi_d}(g, \hat{y}_{1:t-1}, \mathbf{H}),
\end{equation}
where \(\phi_e\) and \(\phi_d\) denote the encoder and decoder parameters, respectively.

Whisper incorporates two primary task-specific tokens: \( \langle| \text{transcribe} |\rangle \) for transcription and \( \langle| \text{translate} |\rangle \) for translation tasks. Additionally, the token \( \langle| \text{notimestamps} |\rangle \) can be used to suppress the decoding of timestamps. The inclusion of language-specific tokens, such as \( \langle| \text{en} |\rangle \) for English, enables Whisper to condition decoding for specific languages and tasks. Furthermore, Whisper supports previous text conditioning by allowing an optional sequence of tokens from prior decoding as input, facilitating context-aware transcription or translation.  

In this study, we utilize the large-v3-turbo variant of Whisper in all our Whisper experiments. This version builds on prior work on distilling Whisper models~\cite{gandhi2023distil} and reduces the number of decoder layers from 32 (large-v3) to 4 without significantly harming the model's performance. This architectural modification significantly reduces autoregressive decoding time, making it more practical for real-world applications. The large-v3-turbo model was trained on a mixture of 1 million hours of weakly labeled audio and 4 million hours of pseudo-labeled audio derived from the large-v3 model. In this work, we turn Whisper into a TS-ASR system by extending and adapting it to condition on diarization information in order to decode speakers of interest.

\subsection{CTC Head for Whisper}
Inspired by~\cite{watanabe2017hybrid}, we incorporate a connectionist temporal classification (CTC)~\cite{graves_06_ctc} head into Whisper to improve the alignment of encoder representations and reduce decoder hallucinations. The CTC head operates on the encoder's hidden representations, enforcing better correspondence between input audio and output tokens. 

It first applies a Transformer layer, followed by two 1-dimensional convolutional layers with a stride of 2, reducing the sequence length from 1500 to 375 (approximately matching the maximum sequence length in the decoder). This subsampling helps to reduce the computational overhead caused by the final linear projection, which maps the hidden representations from dimensionality \( d_m \) to vocabulary size \( V \approx 50k \).

Formally, given the encoder's hidden representations \( \mathbf{H} \in \mathbb{R}^{d_m \times T/2} \), the CTC head computes:
\begin{equation}
    \mathbf{Z} = \text{Linear}(\text{Conv}(\text{SelfAttention}(\mathbf{H}))),
\end{equation}
where \( \mathbf{Z} \in \mathbb{R}^{V \times T/8} \) represents the output logits.

The CTC head enables efficient single-pass decoding and accurate results when combined with the autoregressive decoder~\cite{hori_joint_2017}. It also supports self-speculative decoding~\cite{leviathan_23_speculative}.

\subsection{Joint CTC/Attention Decoding with Whisper}
We observed improved convergence when the CTC head does not ge\-ne\-rate timestamp tokens, which led us to modify the ESPNet~\cite{watanabe2018espnet} CTC prefix scoring implementation\footnote{\url{https://github.com/espnet/espnet/blob/master/espnet/nets/ctc_prefix_score.py}}. Specifically, we adjusted the scoring procedure to retain current states without restoring the next tokens when the autoregressive decoder prefers timestamp tokens. This modification facilitates joint CTC/Attention decoding, wherein the CTC head operates over a distinct subset of labels, diverging from those used by the autoregressive attention-based decoder.

The decoding objective in this setup is defined as a combination of the sequence probabilities from the CTC and attention-based models. Let \( C \) be a sequence, \( p_{\text{ctc}}(C|\mathbf{X}) \) be the sequence probability given by the CTC model, and \( p_{\text{att}}(C|\mathbf{X}) \) be the sequence probability given by the attention-based model. 
The decoding objective is then formulated as:
\begin{equation}
 \label{eq:decoding}
    \hat{C} = \arg \max_{C \in \mathcal{U}^*} \left( \lambda \log p_{\text{ctc}}(C|\mathbf{X}) + (1 - \lambda) \log p_{\text{att}}(C|\mathbf{X}) \right),
\end{equation}
where \( \lambda \) is a weight parameter controlling the balance between the CTC and attention model outputs~\cite{hori_joint_2017} and $\mathcal{U}^*$ is the set of sequences, given by a beam of the top-$k$ most likely hypotheses.

Additionally, we streamlined the implementation by vectorizing key o\-pe\-ra\-tions, improving computational efficiency, and enabling support for batched beam decoding. Our implementation is made publicly available\footnote{\url{https://github.com/BUTSpeechFIT/TS-ASR-Whisper}}.

\section{Diarization-Conditioned Whisper}
\label{sec:dcwv2}
In this section, we introduce the proposed approaches for conditioning on speaker activity. While Whisper serves as the base single-speaker ASR model for our explanations, these methods are generalizable and can be applied to other models, as shown in Section~\ref{sec:non_whisper_models}.  

We begin by defining conditioning masks derived from the speech activity of each speaker, which serves as the foundation for all our proposed systems. We then describe three distinct mechanisms for utilizing these masks:
\begin{enumerate}
    \item \textbf{Input masking}: directly masks the input audio based on speaker activity.
    \item \textbf{Query-key biasing}: selectively biases the attention weights using information from the masks.
    \item \textbf{Frame-level diarization-dependent transformations}: incorporate the full masks to condition encoder representations in a more comprehensive manner.
\end{enumerate}

\subsection{Silence, Target, Non-Target, and Overlap Masks}
\label{sec:stno_mask}

Let $\mathbf{D} \in [0,1]^{S \times T}$, where $S$ is the number of speakers in the recording, and $T$ is the number of frames. 
    The matrix $\mathbf{D}$ represents the diarization output, with $d(s, t)$ denoting the probability that the speaker $s$ is active in time frame $t$. 

The dependency on the number of speakers in $\mathbf{D}$ can be a limiting factor for easily incorporating this mask into the model. 
To address this, we treat each speaker independently. Let $s_k$ represent the target speaker. 
We define a distribution over the following mutually exclusive events for a frame at time $t$:

\begin{itemize}
    \item ${\mathcal{S}}$: The time frame $t$ represents silence.
    \item ${\mathcal{T}}$: The target speaker, $s_k$, is the only active speaker in the time frame $t$.
    \item ${\mathcal{N}}$: One or more non-target speakers, $s \neq s_k$ are active and the target speaker, $s_k$, is not active at the time frame $t$.
    \item ${\mathcal{O}}$: The target speaker $s_k$ is active while at least one non-target speaker $s \neq s_k$ is also active at time frame $t$, denoting an overlap.
\end{itemize}

The probabilities of these events occurring in the time frame $t$ can be calculated as:

\begin{align}
\label{eq:p_S_t}
    p_{\mathcal{S}}^t  &= \prod_{s=1}^S (1 - d(s, t)) \\
    p_{\mathcal{T}}^t  &= d(s_k, t)  \prod_{\substack{s=1 \\ s \neq s_k}}^S (1 - d(s, t)) \\
    p_{\mathcal{N}}^t  &= \left(1 - p_{\mathcal{S}}^t\right) - d\left(s_k, t\right) \\
\label{eq:p_O_t}
    p_{\mathcal{O}}^t  &= d(s_k, t) - p_{\mathcal{T}}^t
\end{align}

This definition allows us to use a fixed-sized STNO (Silence, Target, Non-target, Overlap) mask $\mathbf{M}^t = \begin{bmatrix} p_{\mathcal{S}}^t & p_{\mathcal{T}}^t & p_{\mathcal{N}}^t & p_{\mathcal{O}}^t \end{bmatrix}^{\top}$. 

Note that the mask is speaker-dependent, so decoding each target speaker involves using a different STNO mask, which results in a different transcript.

\subsection{Input Masking}
\label{sec:mask}
Let $\mathbf{x} \in \mathbb{R}^{T}$ denote the input audio signal. The masked signal $\mathbf{x}_{\text{masked}}$ is computed as:

\begin{align}
    \mathbf{x}_{\text{masked}}(t) &= \mathbf{x}(t) \cdot (p_{\mathcal{T}}^t + p_{\mathcal{O}}^t),
\end{align}
Hence, if the target speaker is not active, the output audio signal is equal to 0 (i.e., silence).

However, similar to source separation approaches, this method has li\-mi\-ta\-tions. It can introduce artifacts because it creates an unnatural version of the input signal. Furthermore, it treats all frames in which the target speaker is active equally, without taking into account if there is an overlap with other speakers, which can potentially harm the performance.

\subsection{Query-Key Biasing Conditioning}
\label{sec:qk_bias}

An alternative approach to steer the model's attention away from non-target segments is to integrate target-speaker masks with the model's internal representations by incorporating them into encoder/decoder attention masks.
Compared to the input masking method, attention masking does not introduce artificial silence in the audio signal, reducing the chance of artifacts and potentially leading to better performance.

For simplicity, let us assume a single attention head. Let $\mathbf{W}_q, \mathbf{W}_k \in \mathbb{R}^{{d_m} \times {d_m}}$ denote the query and key projection matrices and $\mathbf{q}_i, \mathbf{k}_j \in \mathbb{R}^{d_m}$ the query and key, respectively. The unnormalized attention score between $\mathbf{q}_i, \mathbf{k}_j$ is defined as:
\begin{equation}
     \label{eq:sdpa_element}
     a_{ij} = \frac{(\mathbf{W}_q \mathbf{q}_i)^T (\mathbf{W}_k \mathbf{k}_j)}{\sqrt{d_m}}.
\end{equation}
To obtain normalized attention weights, the softmax function is applied across the $j$-dimension of $a_{ij}$

If we assume that acoustic information is aligned across time, masking out non-target speaker frames forces the model to ignore information irre\-le\-vant to the target speaker transcript (i.e., other speakers, silence, etc.). However, pure attention masking leaves the model no chance for unmasking and possibly attending to non-target frames, which makes adaptation and speaker-tracking learning impossible.

As a~solution, we decided to bias the encoder self-attention and the decoder cross-attention by extending queries, and keys and initializing corresponding projections in the following way:

\begin{equation}\label{eq:qkb1}
\hat{\mathbf{q}}_{i} = \begin{bmatrix}
\mathbf{q}_i \\
1
\end{bmatrix},
\hat{\mathbf{k}}_{j} = \begin{bmatrix}
\mathbf{k_j} \\
-c
\end{bmatrix},
\hat{\mathbf{W}}_{q} = \begin{bmatrix}
\mathbf{W}_{q} & \textbf{0} \\
\textbf{0} & 1
\end{bmatrix},
\hat{\mathbf{W}}_{k} = \begin{bmatrix}
\mathbf{W}_{k} & \textbf{0} \\
\textbf{0} & 1
\end{bmatrix}
\end{equation}
where $c \in \mathbb{R}^{+}_{0}$ is a bias factor set to  0 if  $k_j$ corresponds to a target speaker frame and to a predefined constant otherwise. We name this approach query-key biasing conditioning (QKb). 

It is easy to observe that, after initialization, if $k_j$ represents a~target speaker frame, $a_{ij}$ remains intact. On the other hand, if $k_j$ represents a~non-target speaker frame, the calculation of $a_{ij}$ changes as:
\begin{equation}
\label{eq:biased_qk_dots}
\hat{\mathbf{W}}_q \hat{\mathbf{q}}_i = \begin{bmatrix}
\mathbf{W}_q \mathbf{q}_i \\
1
\end{bmatrix},\quad
\hat{\mathbf{W}}_k \hat{\mathbf{k}}_j = \begin{bmatrix}
\mathbf{W}_k \mathbf{k}_j \\
-c
\end{bmatrix},
\end{equation}
\begin{equation}
\label{eq:biased_dotproduct}
\begin{bmatrix}
(\mathbf{W}_q \mathbf{q}_i)^T & 1
\end{bmatrix}
\begin{bmatrix}
\mathbf{W}_k \mathbf{k}_j \\
-c
\end{bmatrix}
=
(\mathbf{W}_q \mathbf{q}_i)^T (\mathbf{W}_k \mathbf{k}_j) - c.
\end{equation}
It is important to note that fine-tuning the Whisper model with extended queries and keys changes the extended attention projection matrices $\hat{\mathbf{W}}_{q}$ and $\hat{\mathbf{W}}_{k}$, which controls the level of attention biasing.

\subsubsection{Shifted Positional Embeddings}
Masked silences within an utterance can lead to hallucinations and instabilities during Whisper's training. The main reason is that the decoder cross-attends to discontinuous parts of the encoder embedding sequence after QK biasing is applied. To account for this, instead of adding the original sequence of positional embeddings, we shift the positional embeddings on target speaker frames and repeat the previous positional embeddings on the non-target ones (i.e., for the following sequence of target and non-target frames: TTTNNTT, the positional embeddings would represent the following positions: 1233345), which ensures that the embeddings the decoder attends to have continuous positions.

\subsection{Frame-Level Diarization Dependent Transformations}
\label{sec:fddt}

\begin{figure}
    \centering
    \includegraphics[width=0.8\textwidth]{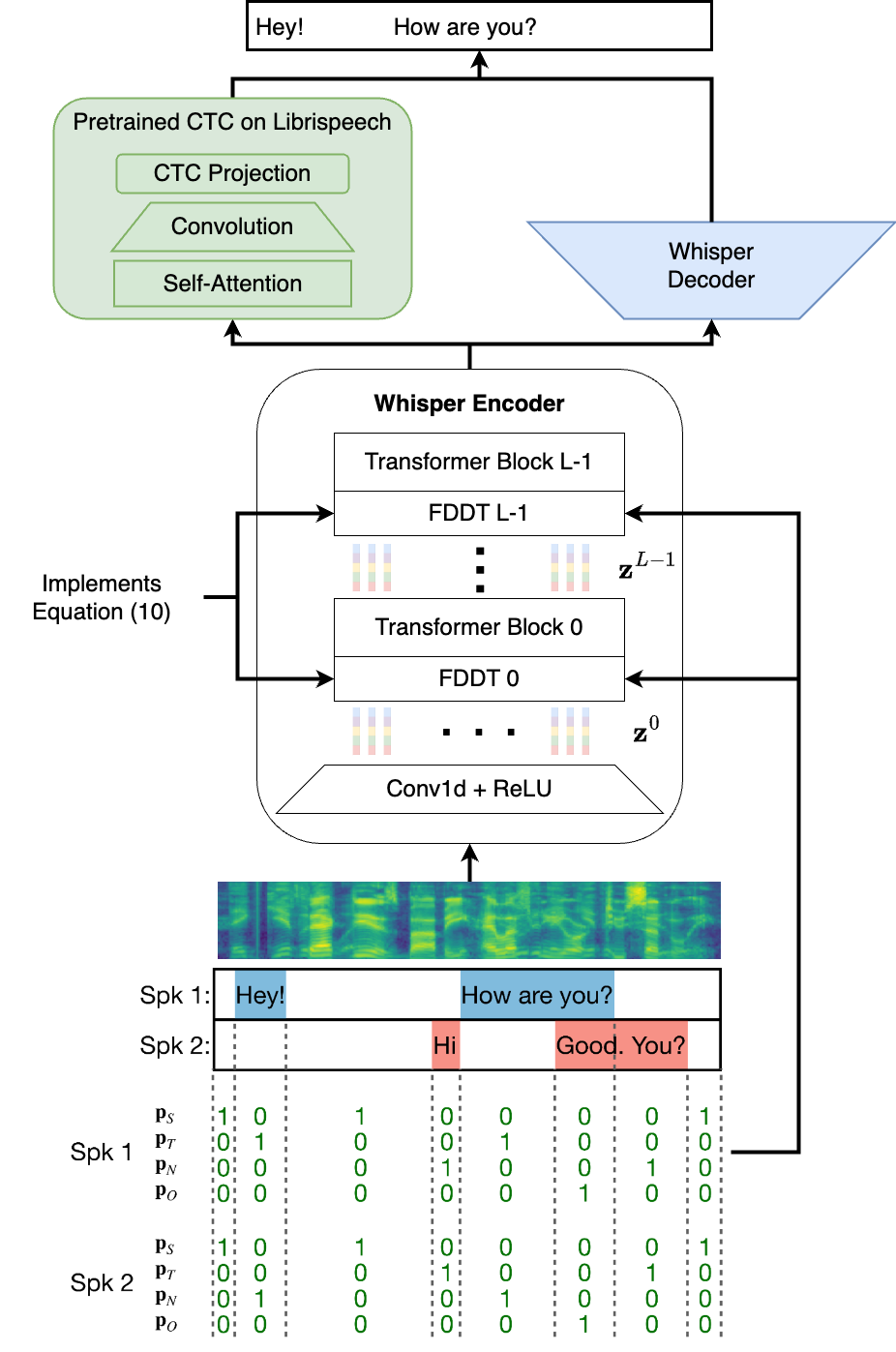}
    \caption{Proposed Diarization-Conditioned Whisper model with STNO mask example.}
    \label{fig:ctc_whisper}
\end{figure}

The third and most sophisticated approach to diarization conditioning is frame-level diarization dependent transformations (FDDT) depicted in Figure~\ref{fig:ctc_whisper}. It modifies the frame-by-frame model inputs based on the diarization outputs and unlike the previous methods, it uses all four STNO masks.

Let $\mathbf{Z}^l \in \mathbb{R}^{d_{{m}} \times T}$ represent the frame-by-frame inputs to the $l$-th (Transformer) layer. We transform these hidden representations by applying four affine STNO layer- and class-specific transformations: $\mathbf{W}_{\mathcal{S}}^l, \mathbf{W}_{\mathcal{T}}^l, \mathbf{W}_{\mathcal{N}}^l, \mathbf{W}_{\mathcal{O}}^l \in \mathbb{R}^{d_{{m}} \times d_{{m}}}$ together with biases $\mathbf{b}_{\mathcal{S}}^l, \mathbf{b}_{\mathcal{T}}^l, \mathbf{b}_{\mathcal{N}}^l, \mathbf{b}_{\mathcal{O}}^l \in \mathbb{R}^{d_{m}}$ to obtain new speaker-specific hidden representations $\hat{\mathbf{Z}}^l = [\hat{\mathbf{z}}^l_1, \ldots, \hat{\mathbf{z}}^l_T]$ as
\begin{align}
\label{eq:FDDT}
\hat{\mathbf{z}}^l_t = &\left( \mathbf{W}_{\mathcal{S}}^l \mathbf{z}^l_t + \mathbf{b}_{\mathcal{S}}^l \right) p^t_{\mathcal{S}} + 
\left( \mathbf{W}_{\mathcal{T}}^l \mathbf{z}^l_t + \mathbf{b}_{\mathcal{T}}^l \right) p^t_{\mathcal{T}}  \nonumber \\
 &+ \left( \mathbf{W}_{\mathcal{N}}^l \mathbf{z}^l_t + \mathbf{b}_{\mathcal{N}}^l\right) p^t_{\mathcal{N}} + 
\left( \mathbf{W}_{\mathcal{O}}^l \mathbf{z}^l_t + \mathbf{b}_{\mathcal{O}}^l \right) p^t_{\mathcal{O}},
\end{align}
In other words, the hidden representations $\mathbf{z}^l_t$ are transformed using a convex combination of the four STNO class-specific affine transformations, weighted by the corresponding STNO class probabilities~\eqref{eq:p_S_t}-\eqref{eq:p_O_t}. When using a hard STNO mask (i.e., one-hot encoding), equation \eqref{eq:FDDT} simplifies selecting and applying one of the four class-specific transformations for each frame. Note that the same transformation is applied to all frames with identical STNO masks.

The matrices $\mathbf{W}_{\mathcal{S}}^l, \mathbf{W}_{\mathcal{T}}^l, \mathbf{W}_{\mathcal{N}}^l, \mathbf{W}_{\mathcal{O}}^l$ are designed to transform the hidden representations into a space where speaker distinction is more effective, or where certain components of the signal can be suppressed. These transformations are essential for the model to correctly identify and isolate the target speaker while handling other sources of noise.

However, fine-tuning the model with randomly initialized FDDT matrices could disrupt its already learned internal representations~\cite{polok2024targetspeakerasrwhisper}. To mitigate this risk, we employ a suppressive initialization strategy. In this strategy, we initialize $\mathbf{W}_{\mathcal{S}}^0$ and $\mathbf{W}_{\mathcal{N}}^0$ as zero matrices, which effectively suppresses the influence of other speakers; while the other parameters $\mathbf{b}_{\mathcal{S}}^l, \mathbf{b}_{\mathcal{T}}^l, \mathbf{b}_{\mathcal{N}}^l, \mathbf{b}_{\mathcal{O}}^l$ (set to zero vectors), and $\mathbf{W}_{\mathcal{T}}^l, \mathbf{W}_{\mathcal{O}}^l$ (set to identity matrices) are initialized to maintain stability and allow the model to process the STNO classes effectively.

\subsection{Co-Attention Module for Speaker Interaction in TS-ASR}
\begin{figure}
    \centering
    \includegraphics[width=0.8\linewidth]{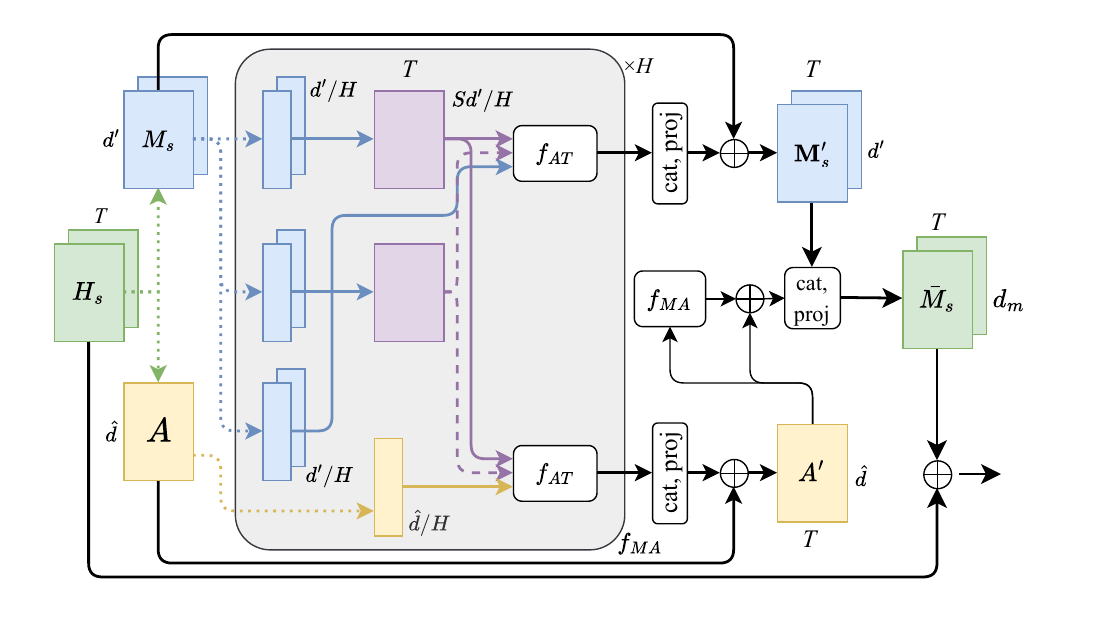}
    \caption{Scheme of Co-Attention module. Dotted lines depict affine transformation followed by layer norm. The gray rectangle depicts the definition of multi-headed attention derived in \eqref{eq:mha}. Inputs to scaled-dot product $\att$ attention are queries, keys and values from top to bottom.}
    \label{fig:co_attention}
\end{figure}
So far, we have discussed diarization conditioning approaches for TS-ASR. In the previous sections, we designed the system to process each speaker independently as a target speaker, applying parallel processing for each one. While effective in certain scenarios, this approach can be limiting when transcribing conversations involving multiple speakers, such as in meetings. It is inherently suboptimal as it does not account for interactions between spea\-kers. To address this limitation, we explore methods to model inter-speaker dependencies.

One such method is the Co-Attention module, which we present as a means to effectively capture and manage these dependencies within the context of TS-ASR. Originally proposed for speaker diarization~\cite{horiguchi22_coattention} and later extended to speaker identification tasks~\cite{mosner24_interspeech}, the Co-Attention module processes encoder outputs from multiple speakers, contextualizing these representations while addressing redundancies in overlapping speech transcripts. The architecture of the Co-Attention module is depicted in Figure~\ref{fig:co_attention}.

In the following method description, we adopt the notation introduced in~\cite{mosner24_interspeech}.
Let 
\begin{equation}\label{eq:mha}
    \mha(\mb{X}, \mb{Y}; \theta_\mathrm{MA}) := \W_{O} \left[ \cat{h=1}{H} \att \left( \W_Q^{(h)}\mb{X}, \W_K^{(h)}\mb{X}, \W_V^{(h)}\mb{Y}  \right) \right]
\end{equation} be H-head attention parametrized by $\theta_\mathrm{MA}$,
where $\att(\mb{Q}, \mb{K}, \mb{V})$ denotes scaled dot-product attention~\cite{vaswani2017attention}, $||$ represents concatenation along the first dimension, and $\theta_\mathrm{MA}$ consists of the weight matrices $\{\W_Q^{(h)}, \W_K^{(h)}, \W_V^{(h)} \}_{h=1}^H \cup \{ \W_O \}$.

Our Co-Attention version performs summarization of all speakers' speech, concatenation with per-speaker representations, and enriches original input embeddings with derived contextual information.
Given the encoder outputs $\{\H_s\}_{s=1}^{S}$ for $S$ speakers, where $\H_s \in \bR^{d_m\times T}$, speaker summarization is derived as
\begin{equation}
\A = \LN \left( \W_{A} \left[ \frac{1}{S} \sum_{s} \H_s \right] \right),
\end{equation}
where $\W_{A} \in \bR^{\hat{d} \times d_{m}}$ projects the aggregated summary embeddings to a lower-dimension $\hat{d}$. Per-speaker representations are defined as
\begin{equation}
\mb{M}_s = \LN \left( \W_M \H_s \right) \in \bR^{d' \times T},
\end{equation}
and concatenated across speakers to obtain
\begin{equation}
\mb{M} = \cat{s=1}{S} \mb{M}_s \in \bR^{S d' \times T}.
\end{equation}

Next,  Co-Attention is performed to model interactions between speakers
\begin{equation}
\begin{split}
    \mb{M}_s^{\prime} &= \LN \left( \mha(\mb{M}, \mb{M}_s; \theta) + \M_s \right), \\
    \mb{A}^{\prime} &= \LN \left( \mha(\mb{M}, \mb{A}; \xi) + \A \right),
\end{split}
\end{equation}
where H-head attention parameters $\theta$ and $\xi$ share the same query and key transformation matrices, $\mb{W}_Q^{(h)}$ and $\mb{W}_K^{(h)}$, resulting in identical attention matrices. Here, $\mb{W}_Q^{(h)}$ and $\mb{W}_K^{(h)}$ are block-diagonal matrices composed of $S$ individual matrices with dimensionality $\bR^{d'/H \times d'}$ all sharing the same parameters.

To further refine the representations, we apply \textit{self-attention} to the contextualized summary embeddings.
\begin{equation}
\bar{\A} = \LN \left( \mha( \mb{A}^{\prime}, \mb{A}^{\prime}; \omega ) + \mb{A}^{\prime} \right).
\end{equation}
Finally, $\bar{\A}$ is concatenated with each speaker's co-attended representation $\mb{M}_s^{\prime}$, and a linear projection maps the combined output back to the original space $\bR^{d_m}$
\begin{equation}
\bar{\M}_s = \W_F \left( \mb{M}_s^{\prime} || \bar{\A} \right), \quad \W_F \in \bR^{d_{m} \times (d' + \hat{d})}.
\end{equation}
The updated representation for speaker $s$ is given by
\begin{equation}
\hat{\H}_s = \H_s + \bar{\M}_s.
\end{equation}

\section{Experimental Setup}
\label{sec:experiments_setup}
This section describes details of our experimental setup, including datasets, evaluation metrics, training procedure, hyperparameters, and lastly, the diarization system.
All our models were implemented in HuggingFace Transformers library~\cite{wolf-etal-2020-transformers}. We used Whisper-large-v3-turbo as it is a faster version of large-v3, and we observed almost no performance degradation.

\subsection{Evaluation Datasets}
To validate the proposed method, we utilized two publicly available multi-speaker datasets: AMI~\cite{Mccowan2005_ami} and NOTSOFAR-1~\cite{vinnikov24_interspeech}. Both have realistic interactions in a far-field setting, presenting English-spoken meetings in a challenging scenario. Statistics about the sets can be found in Tables~\ref{tab:datasets_information1}\,and\,\ref{tab:datasets_information2}.

To be able to compare with other existing methods, we also utilized Libri2Mix~\cite{Cosentino2020LibriMixAO} and LibriCSS~\cite{chen2020continuous}. 

Given the current state of technology and the availability of public datasets, we believe results should be reported on real and not synthetic datasets.

\begin{table*}[t]
    \caption{Numbers of files, minimum and maximum numbers of speakers per recording, and numbers of hours per partition.}
    \label{tab:datasets_information1}
    \setlength{\tabcolsep}{3pt} 
    \centering
    \begin{tabular}{l|ccc|ccc|ccc}
    \toprule
    \multirow{2}{*}{Dataset} & \multicolumn{3}{c|}{train} & \multicolumn{3}{c|}{development} & \multicolumn{3}{c}{test} \\ 
    & \#files & \#spk & \#\,h & \#files & \#spk & \#\,h & \#files & \#spk & \#\,h \\
    \midrule
   AMI & 136 & 3-5 & 80.67 & 18 & 4 & 9.67 & 16 & 3-4 & 9.06 \\
   NOTSOFAR-1 & 526 & 4-8 & 54.27 & 117 & 4-6 & 12.17 & 160 & 3-7 & 16.67 \\
   \midrule
   Libri2Mix & 13900 & 2 & 56.37 & 3000 & 2 & 7.6 & 3000 & 2 & 7.01 \\
   LibriCSS & - & - & - & 7 & 8 & 1.0 & 55 & 8 & 9.09 \\
    \bottomrule
  \end{tabular}
\end{table*}

\begin{table}[t]
    \caption{Information about the percentage of silence, speech with a single speaker, and overlap for each set.}
    \label{tab:datasets_information2}
    \setlength{\tabcolsep}{3.5pt} 
    \centering
    \begin{tabular}{l|ccc|ccc|ccc}
    \toprule
    \multirow{2}{*}{Dataset} & \multicolumn{3}{c|}{train} & \multicolumn{3}{c|}{development} & \multicolumn{3}{c}{test} \\ 
    & \%sil & \%1-spk & \%ov & \%sil & \%1-spk & \%ov & \%sil & \%1-spk & \%ov \\
    \midrule
   AMI & 16.5 & 72.3 & 11.2 & 22.0 & 61.1 & 16.9 & 14.7 & 67.9 & 17.4 \\
   NOTSOFAR-1 & 8.1 & 65.4 & 26.5 & 17.7 & 68.8 & 13.7 & 8.0 & 66.6 & 25.2 \\
   \midrule
   Libri2Mix & 5.5 & 33.9 & 60.6 & 8.4 & 42.6 & 49.0 & 8.1 & 42.8 & 49.1 \\
   LibriCSS & - & - & - & 6.2 & 84.2 & 9.6 & 6.7 & 83.7 & 9.6 \\
    \bottomrule
  \end{tabular}
\end{table}

\subsection{Evaluation Metrics}

Word error rate (WER) is normally used to evaluate single-speaker ASR. It calculates the error of a hypothesis with respect to the reference as the sum of substitutions, insertions, and deletions over the number of words in the reference annotation. However, with recordings having more than one speaker, different recognizers are usually evaluated with different metrics depending on how the hypotheses and references are mapped. In order to be able to compare with relevant previous works, we considered:

\begin{itemize}
    \item Concatenated minimum-permutation WER (cpWER), where for each speaker, all their utterances are concatenated, and the best permutation between hypothesis and reference is used to calculate the standard WER. This metric calculates speaker-attributed ASR errors.
    \item Time-constrained minimum-permutation WER (tcpWER), where the evaluation is like cp-WER but also considering the temporal alignments of the words.
    \item Optimal reference combination WER (ORC-WER), which does not consider the speaker labels and can be used to evaluate speaker-agnostic systems.
    \item Time-constrained optimal reference combination WER (tcORC-WER), where the evaluation optimally matches hypotheses and references without considering speaker labels but ensures that temporal alignments are respected. 
\end{itemize}

To avoid cluttering tables with notation, we do not write `WER' in the headers below.
For a more thorough analysis and comparison of the metrics, we refer the reader to~\cite{Neumann2023MeetEval}.
For time-constrained metrics, we use a collar of 5s.
The diarization error rate (DER), used to evaluate the diarization system, is calculated using collar 0s.

\subsection{Training Details}
\label{sec:training_details}
The training is divided into three consecutive phases: 
\begin{enumerate}
    \item \textbf{CTC preheat} - pre-train only the CTC-related parameters on LibriSpeech 960h~\cite{librispeech} with the rest of the model being frozen. CTC is trained without timestamps as we believe that timestamp prediction should not occur in the encoder and can rather be heuristically derived from the CTC frame-by-frame predictions.
    \item \textbf{FDDT preheat} - pre-train CTC and FDDT-related parameters on the target multi-speaker dataset. We train FDDT parameters with 100$\times$ higher learning rate than the rest (i.e., $2\times10^{-5}$) to improve the model convergence.
    \item \textbf{Full fine-tuning} - fine-tune all the parameters on the target multi-speaker dataset until convergence.
\end{enumerate}

We train all the models with an overall batch size of 64 samples using AdamW~\cite{loshchilov2018decoupled} optimizer with weight decay $1\times10^{-6}$. We warm up the learning rate for 5k steps (i.e., 5k per batch model updates) and then use a linear decay scheduler for the rest of the training. The peak learning rate is set to $2\times10^{-7}$. The CTC loss weight is set to $\lambda=0.3$~\cite{watanabe2017hybrid}. We evaluate the model on the development set at intervals of \(\min(1\ \text{epoch}, 500\ \text{steps})\), monitor the development tcpWER, and use early stopping with a patience of 5 evaluation steps. The maximum number of training steps is set to 50k, and we will select the final model based on the best development tcpWER.
Please note that CTC Preheat training is independent of FDDT preheat and full fine-tuning. For CTC Preheat, we use only 1k warm-up steps with a peak learning rate of \(2\times10^{-4}\), monitoring WER on Librispeech dev-clean and dev-other sets. Other hyperparameters, including the batch size and optimizer settings, remain unchanged.

Furthermore, we do not enforce true casing or lower casing. Instead, we compute the cross-entropy loss with both lowercase and uppercase labels and select the smaller loss value.
We empirically chose the initial value of the QK biasing constant $c$ \eqref{eq:qkb1}-\eqref{eq:biased_dotproduct} to be 50 as we observed that higher values result in fewer hallucinations at the beginning of training, while also making it easier for the model to approximate a hard attention mask by producing very low attention scores without having to scale up the query and key projection weights.

\subsection{Diarization System}
\label{sec:diarization}
We utilized DiariZen~\cite{han2024leveraging}\footnote{\url{https://github.com/BUTSpeechFIT/DiariZen}} which is based on a framework with local end-to-end neural diarization (EEND) followed by speaker embedding clustering~\cite{kinoshita2021integrating,kinoshita2021advances} using pyannote~\cite{bredin2023pyannote,plaquet23_interspeech}. The EEND module combines WavLM~\cite{chen2022wavlm} with Conformer~\cite{gulati2020conformer} layers and is trained using the powerset loss~\cite{plaquet23_interspeech}. The EEND operates on 8-second-long overlapping segments, and for each speaker found in the segment, a speaker embedding is extracted using a ResNet34-based embedding extractor~\cite{wang2023wespeaker}. These are, in turn, clustered by spectral clustering \cite{park2019auto} to obtain the inter-segment mapping between speakers and produce a single output for each recording. 
The hard decisions made by the diarization system are used to provide the corresponding inputs for the following ASR system. Note that we use the same model on all evaluation sets. This model is compliant with the CHiME-8 challenge rules~\cite{vinnikov24_interspeech} for the NOTSOFAR-1 track. Fine-tuning the model on each specific dataset could lead to further improvements.

\section{Experiments}
\label{sec:experiments}

This section presents a comparison of the proposed methods. First, we compare our best configuration with other works considering both ground-truth and system diarization. We use cpWER to assess the quality of speaker-attributed transcripts and ORC-WER to assess the quality of transcripts disregarding speaker assignment errors. However, these metrics do not penalize the model for incorrectly determining the utterance timing, which is one of the qualities that TS-ASR systems should, in our opinion, have. Hence, we also report time-constrained metrics (i.e., tcpWER and tcORC-WER) similar to the NOTSOFAR-1 challenge. Different works have reported different evaluation metrics for different datasets, so we report results using the metrics used in the literature before and also all the time-constrained metrics. We encourage researchers to compare their future work based on time-constrained metrics. Note that calculating ORC-WER for some results is infeasible, so we approximate these values using tcORC-WER instead by increasing the time collar until tcORC-WER converges (the obtained metric cannot be lower than ORC-WER).

First, we focus on comparing the proposed variants of diarization conditioning, presenting results across all datasets using ground-truth diarization, followed by an analysis of the impact of real diarization on the results.
Next, we analyze the effect of joint CTC training and decoding on the NOTSOFAR-1 dataset. We also report the impact of target-speaker training on single-speaker evaluation datasets.
Finally, we show the proposed diarization conditioning when using a different architecture to showcase that the method can be applied to other ASR models.

\subsection{Baseline Comparison}
In this section, we begin by analyzing the performance of our best-performing models, as summarized in Table~\ref{tab:SOTA_comp}, and comparing them to state-of-the-art results reported in the literature. Specifically, we use the best-performing variants of our approach: the multi-domain (MD) FDDT model for AMI-sdm, NOTSOFAR-1, and LibriCSS, and the single-domain (SD) fine-tuned version with Co-Attention for Libri2Mix. These configurations were selected to highlight our most effective methods on each dataset. For a more detailed comparison of all proposed methods, refer to Table~\ref{tab:FDDT_full}.

Even though comparisons across datasets are inherently challenging, as different studies report results on different datasets, our models demonstrate consistently strong performance. On AMI-sdm, to the best of our know\-ledge, we achieve state-of-the-art results with ground-truth diarization. Notably, the degradation when switching to real diarization is relatively small for ORC-WER. However, for cpWER, the impact is significantly larger, indicating that errors related to speaker label assignment (linked to the confusion errors in DER) have a more substantial effect than inaccuracies in segment boundaries.

Our proposed method also achieves the best results to date on the Libri2\-Mix and LibriCSS datasets for both real- and ground-truth diarization. However, while these results are noteworthy, we consider evaluations on artificial datasets like Libri2Mix to be of limited practical relevance.

\begin{table}[ht]
    \centering
    \caption{%
Comparison of the proposed system alongside various multi-speaker ASR systems. The top section includes systems where no additional information about speaker identity or segmentation is provided. Our results are obtained using a diarization system. The bottom section features models that directly or indirectly utilize ground-truth (oracle) diarization information. Proposed ORC\,WER results marked with $\star$ were approximated by increasing the collar for tcORC\,WER. Results marked with $\dagger$ are evaluated on utterance groups, where the model is not penalized for speaker confusions outside the window of the utterance group, which significantly reduces cpWER.  
    }\label{tab:SOTA_comp}
    \setlength{\tabcolsep}{2pt} 
        \small{\begin{tabular}{lcc|cc|cccc|cc}
        \toprule
         & \multicolumn{2}{c|}{AMI-sdm}  & \multicolumn{2}{c|}{NOTSOFAR-1}   & \multicolumn{4}{c|}{Libri2Mix} & \multicolumn{2}{c}{LibriCSS}  \\
         & \multicolumn{2}{c|}{test} & \multicolumn{2}{c|}{eval-small}  & \multicolumn{2}{c}{test-clean} & \multicolumn{2}{c|}{test-both} & \multicolumn{2}{c}{test}  \\
         &  \rotatebox[origin=c]{90}{cp} & \rotatebox[origin=c]{90}{ORC}  & \rotatebox[origin=c]{90}{tcp} & \rotatebox[origin=c]{90}{tcORC}  & \rotatebox[origin=c]{90}{cp} & \rotatebox[origin=c]{90}{ORC}  & \rotatebox[origin=c]{90}{cp} & \rotatebox[origin=c]{90}{ORC}  & \rotatebox[origin=c]{90}{cp} & \rotatebox[origin=c]{90}{ORC} \\
        \midrule
        Kanda et al.~\cite{Kanda2021MultiTalkerASR}
        &  $21.2^{\dagger}$    &      &      &      &      &      &      &      &  \\
        Raj et al.~\cite{raj23_surt2}
        &      & $44.6^{\dagger}$ &      &      &      &      &      &      &     & 16.9  \\
        Zarandi et al.~\cite{zarandi_23_chubert}
        &      &      &      &      &      & 7.8  &      &      &  \\
        Vinnikov et al.~\cite{vinnikov24_interspeech}
        &      &      & 41.4 & 35.5 &      &      &      &  \\
        Cornell et al.~\cite{cornell2024one}
        & 24.5 &      &      &      &      &      &      &      &  \\
        Niu et al.~\cite{niu24_chime}
        &     &      & 22.2 & 17.7 &      &      &      &      &  \\
        Ours (real diar.) 
        & 23.6 & $18.0^\star$ & 33.5 & 22.6 & 6.0 & 6.0 & 15.0 & 14.9 & 8.5 & $6.5^\star$\\
        \midrule
        Cornell et al.~\cite{cornell2024one}
        & 21.1 &      &      &      &      &      &      &      &  \\
        Ma et al.~\cite{ma2024extending} 
        &      &      &      &      & 12.0 &      & 26.4 &      &  \\
        Zhang et al.~\cite{Zhang2023}
        &      &      &      &      &      &      & 23.5 &      &  \\
        Meng et al.~\cite{meng24c_interspeech}
        &      &      &      &      & 4.7  &      &      &      &  \\
        Guo et al.~\cite{guo2024sq}
        & 22.0 &      &      &      &      &      & 14.6 &      &  \\
        Ours (oracle diar.)    
        & 17.2 & 
        $16.5^\star$ 
        & 19.7 & 19.1 & 4.4  & 4.4  & 10.9 & 10.9 & 5.6 &  
        $5.5^\star$\\
        
        \bottomrule
    \end{tabular}}
\end{table}

\subsection{Input masking vs. QK biasing vs. FDDT}
Table~\ref{tab:FDDT_full} presents the performance of the methods proposed in Section~\ref{sec:dcwv2}. It can be seen that the out-of-the-box Whisper model does not perform well, as it lacks a mechanism to prevent transcribing all present speech, irrespective of who is considered the target speaker. By masking the non-target speaker audio (Input masking), we improve on all datasets substantially.

Furthermore, we can see that QK biasing after initialization (and before fine-tuning) does not perform well, reaching WER metrics above 100\%, suggesting strong levels of hallucination. After fine-tuning, we can see that not shifting positional embeddings results in better performance with regard to both tcpWER and tcORC WER. We further analyze the difference between these two modifications in Table~\ref{tab:QK_shift_tab}.

Lastly, the third section of Table~\ref{tab:FDDT_full} presents FDDT. First, it can be observed that the method performs comparably to input masking right after suppressive initialization, suggesting that the initialization does not break the original Whisper model compared to QK biasing. Furthermore, after single-domain (SD) fine-tuning (i.e., fine-tuning the model only on the corresponding training set), we can observe a massive improvement on all the datasets.

Our model faces a challenge with fully overlapped speech, such as in Libri2Mix, where two speakers spew concurrently most of the time. This limitation arises from the model design, where each target speaker is decoded by an independent instance of the TS-ASR model. With fully overlapped speech, it might be difficult for an independent TS-ASR instance to determine which speaker's speech is responsible for decoding. As a consequence, multiple TS-ASR instances can decide to decode speech from the same speaker. To mitigate this, we incorporate a Co-Attention mechanism that allows the model to compare information across target speaker channels (i.e., TS-ASR instances). This mechanism allows the channels to collaboratively decide which instance is responsible for decoding each speaker in the input utterance. By resolving this ambiguity, Co-Attention reduces errors on Libri2Mix by 2–3\% absolute.

However, the improvement is not as significant on AMI or NOTSOFAR-1, as these datasets have a significantly lower percentage of overlapped speech, and also, more speakers are present, which makes the scenario more challenging. 

FDDT MD refers to a multi-domain model, which utilizes training data from AMI-sdm, NOTSOFAR-1, and Libri2Mix weighted with a ratio of 4:4:1. We selected the best-performing checkpoint based on the NOTSOFAR-1 development set.

It can be seen that the model outperforms the other approaches on both real datasets. However, it performs worse on Libri2Mix, suggesting the domain mismatch between real-world and synthetic mixtures. It also demonstrates that the increased amount of data is not solving the full overlap issue and that a speaker interaction module is indeed necessary.

The addition of Co-Attention to the MD FDDT model slightly improves performance on Libri2Mix compared to MD FDDT alone, indicating its potential to better handle fully overlapped speech in synthetic mixtures.

However, it remains inferior to the performance demonstrated by SD + Co-Attention on Libri2Mix, where the smaller, more targeted domain training likely aligns better with the dataset's specific characteristics. On the other hand, MD FDDT + Co-Attention does not offer improvements on AMI or NOTSOFAR-1, which may be attributed to the variability in the number of speakers and the dynamic interaction patterns in these real-world datasets. This variability affects the normalization of attention scores within the Co-Attention module. This suggests that while Co-Attention provides some benefits in specific scenarios, its integration with MD FDDT may require further design changes to handle diverse speaker configurations more effectively.

\begin{figure}[ht]
    \centering
    \includegraphics[width=0.8\textwidth]{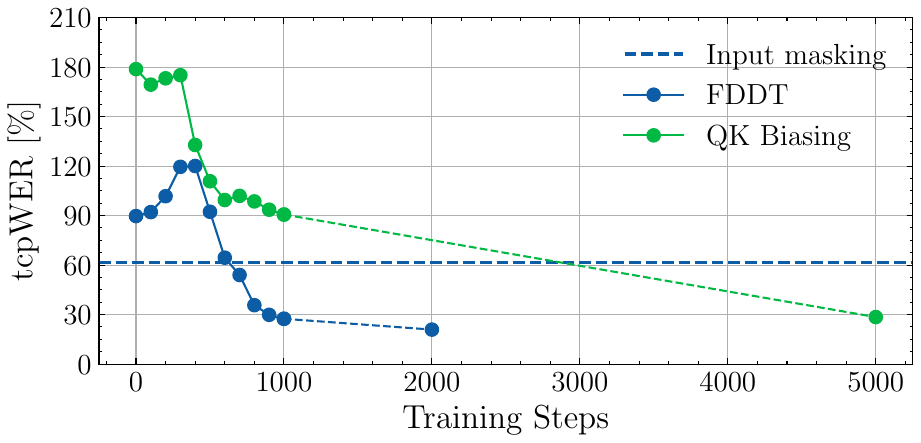}
    \caption{Test TCP-WER as a function of training steps for the NOTSOFAR-1 model evaluated on the eval-small dataset.}
    \label{fig:tcp-wer}
\end{figure}

Furthermore, Figure~\ref{fig:tcp-wer} shows the evolution of tcpWER of QK biasing and FDDT. It can be seen that FDDT converges much quicker than QK biasing, which is mainly caused by the non-disturbing initialization. After 1000 steps, the FDDT approach reaches tcpWER below 30\,\%, suggesting that FDDT is an effective module that quickly turns Whisper into a target-speaker model.

\begin{table}[ht]
    \centering
    \caption{%
Comparison of different diarization-conditioning methods. The table is divided into three sections: the first section presents the performance of vanilla Whisper and Input Masking methods, which do not require additional training; the second and third sections show the results for QK biasing and FDDT with different configurations. An asterisk (*) indicates values that could not be computed due to scalability issues, while a dash (–-) denotes single-domain setups (LibriCSS) where training data is unavailable.
    }\label{tab:FDDT_full}
    \setlength{\tabcolsep}{2pt} 
        \small{\begin{tabular}{lcc|cc|cccc|cc}
        \toprule
         & \multicolumn{2}{c|}{AMI-sdm}  & \multicolumn{2}{c|}{NOTSOFAR-1}   & \multicolumn{4}{c|}{Libri2Mix} & \multicolumn{2}{c}{LibriCSS}  \\
         & \multicolumn{2}{c|}{test} & \multicolumn{2}{c|}{eval-small}  & \multicolumn{2}{c}{test-clean} & \multicolumn{2}{c|}{test-both} & \multicolumn{2}{c}{test}  \\
         &  \rotatebox[origin=c]{90}{tcp} & \rotatebox[origin=c]{90}{tcORC}  & \rotatebox[origin=c]{90}{tcp} & \rotatebox[origin=c]{90}{tcORC}  & \rotatebox[origin=c]{90}{tcp} & \rotatebox[origin=c]{90}{tcORC}  & \rotatebox[origin=c]{90}{tcp} & \rotatebox[origin=c]{90}{tcORC}  & \rotatebox[origin=c]{90}{tcp} & \rotatebox[origin=c]{90}{tcORC} \\
        \midrule
        Whisper
        & 220.0& 212.0& 260.1& *    & 66.1 & 66.1 & 69.4 & 69.4 & 588.2 & *  \\
        Input masking   
        & 52.8 & 47.9 & 61.6 & 54.0    & 42.2 & 42.1 & 47.9 & 47.9 & 56.2 & 55.0  \\
        \midrule
        QKb init.
        & 276.3 & * & 260.0 & * & 323.6 & * & 339.4  & * & 990.9 & *  \\
        QKb w shift
        & 55.8 & 54.1 & 65.3 & * & 9.9 & 9.9 & 16.5  & 16.5 & -- & --  \\
        QKb w/o shift
          & 47.8  & 46.6     & 28.2 & * & 7.9 & 7.9 & 16.4 & 16.4 & -- & --  \\
        \midrule
        FDDT init.  
        & 78.3 & 68.7 & 89.7 & 77.5 & 100.6& 96.1 & 105.9& 100.4& 102.0 & 101.9 \\
        FDDT SD      
        & 17.8 & 17.5 & 20.9 & 20.3 & 6.3  & 6.3  & 13.8 & 13.8 & -- & -- \\
        + CoAttention     
        &  \textbf{17.5}  &  17.2  & 20.8 & 20.3 & \textbf{4.4}  & \textbf{4.4}  & \textbf{11.0} & \textbf{11.0} & -- & -- \\
        FDDT MD  
        & 17.6 & \textbf{16.7} & \textbf{19.7} & \textbf{19.1} & 6.9  & 6.9  & 15.9 & 15.9 & \textbf{8.8} & \textbf{8.8}\\
        + CoAttention 
        & 18.1 & 17.7 & 20.0 & 19.4 & 5.8 & 5.8 & 14.4 & 14.4 & 11.0 & 11.0 \\
        
        
        \bottomrule
    \end{tabular}}
\end{table}

\begin{table}[ht]
    \centering
    \caption{Comparison of cpWER and tcpWER between QK biasing with and without shifted positional embeddings with ground-truth diarization.}
    \label{tab:QK_shift_tab}
    \setlength{\tabcolsep}{6pt} 
        \small{\begin{tabular}{lcc|cc|cccc}
        \toprule
         & \multicolumn{2}{c|}{AMI-sdm}  & \multicolumn{2}{c|}{NOTSOFAR-1}   & \multicolumn{4}{c}{Libri2Mix} \\
         & \multicolumn{2}{c|}{test} & \multicolumn{2}{c|}{eval-small}  & \multicolumn{2}{c}{test-clean} & \multicolumn{2}{c}{test-both} \\
         &  \rotatebox[origin=c]{90}{cp} & \rotatebox[origin=c]{90}{tcp}  & \rotatebox[origin=c]{90}{cp} & \rotatebox[origin=c]{90}{tcp}  & \rotatebox[origin=c]{90}{cp} & \rotatebox[origin=c]{90}{tcp}  & \rotatebox[origin=c]{90}{cp} & \rotatebox[origin=c]{90}{tcp} \\
        \midrule

        QKb w shift
        & 21.3 & 55.8 & 25.2 & 65.4 & 9.9 & 9.9 & 16.5 & 16.5 \\
        QKb w/o shift
        & 45.9     &   47.8   & 27.3 & 28.5 & 7.9 & 7.9 & 16.4 & 16.4 \\
        \bottomrule
    \end{tabular}}
\end{table}

\subsubsection{QK Biasing Shift}
Table \ref{tab:QK_shift_tab} shows differences in cpWER and tcpWER between QK biasing with and without shifted positional embeddings on three datasets: AMI-sdm, NOTSOFAR, and Libri2Mix. An immediate observation is that both metrics on Libri2Mix are exactly the same compared to the real datasets, which is mainly caused by Libri2Mix being a short-form dataset --- i.e., utterances are shorter than 30s; thus, Whisper does not utilize timestamps for inference. Also, the majority of the words are going to be within the collar boundaries, not influencing reference alignment.

Furthermore, while QK biasing with shift provides a lower cpWER on real datasets, it is worse on Libri2Mix, also probably due to the short-form and synthetic nature of this dataset.

This highlights the risk of drawing incorrect conclusions from interpretations developed on artificial evaluation data. Therefore, our analyses and hypotheses are based on results obtained from the real datasets AMI-sdm and NOTSOFAR-1. 

It can be seen that shifting positional embeddings helps transcription quality on AMI-sdm (24.6\% absolute cpWER improvement) and NOTSOFAR (2.1\% absolute cpWER improvement) datasets. We suspect that encoder attention masking creates ``holes" in the positional information in the encoder's output to which the decoder cross-attends, potentially causing the decoder to lose track of the position in the input sequence and hallucinate.

On the other hand, the gap between cp and tcp WER suggests that the positional information is not accurate after shifting positional embeddings and that the Whisper model timestamp prediction heavily relies on the positional embeddings added to the encoder input. However, if we do not shift the positional embeddings, the gap between cp and tcpWER decreases to a few percent on both the AMI-sdm and NOTSOFAR-1 datasets.

\subsection{Performance with Real Diarization}
We have demonstrated that our proposed DiCoW model achieves strong performance when high-quality speaker and segmentation information is provided. We also evaluated our proposed system with real diarization (cf. Section~\ref{sec:diarization}). In this section, we focus on a more detailed analysis of the impact of diarization errors on DiCoW. We use the FDDT model fine-tuned on all datasets (AMI-sdm, NOTSOFAR-1, Libri2Mix).

As shown in Table~\ref{tab:real_diarization}, the performance deteriorates significantly with respect to the system with ground-truth diarization. This decline is primarily attributed to errors introduced by the diarization process (cf.~Table\,\ref{tab:DER_performance}).

On the NOTSOFAR-1 eval-small dataset, we achieved a tcpWER of 33.5\,\% (22.6\,\%~tcORC-WER), with 14.5\,\% (8.3\,\%) deletions, 10\,\% (3.6\,\%) insertions, and 9\,\% (10.7\,\%) substitutions, indicating that the system struggles with omissions in this scenario. This is expected, given that this dataset features recordings with a higher number of speakers on average and significantly more overlapping speech, both of which increase the likelihood of confusion errors during diarization. In contrast, LibriCSS has fewer overlapping segments and fewer speakers per recording. However, most of the diarization errors in LibriCSS (cf. Table~\ref{tab:DER_performance}) are caused by missed speech, which results in 4\,\% deletions, 4\,\% substitutions and 3\,\% insertions.

The errors are partly due to the model being trained with ground-truth diarization, which provides accurate speaker boundaries and correctly labeled silence. In contrast, real diarization can miss portions of a speaker's speech, particularly in scenarios with more than two overlapping speakers. The system struggles to recover from missed speech segments because it has not encountered such cases during training. Moreover, we initialize the FDDT parameters in such a way that the model ignores frames marked as silence from the very beginning (cf.~Section\,\ref{sec:fddt}). This further limits the system's ability to recover from diarization errors.  

The proposed DiCoW model is designed to handle soft diarization decisions in the form of frame-by-frame speaker activity probabilities $d(s, t)$. However, in this work, we use only hard diarization decisions for both training and inference. As the result, the probabilities $d(s, t)$ and, consequently, the STNO class probabilities~\eqref{eq:p_S_t}-\eqref{eq:p_O_t} are restricted to binary values $0$ or $1$.

For training, we rely solely on hard ground-truth diarization decisions. Using real diarization outputs introduces the additional challenge of aligning the speakers identified by the diarization system with the ground-truth speakers labeled in the ASR annotations. This alignment becomes especially problematic when the diarization system predicts an incorrect number of speakers.

In our system submitted to the CHiME-8 challenge~\cite{polok24_butjhu} (closely following the method described in this paper), we demonstrated that soft diarization decisions can improve ASR decoding performance. However, subsequent analysis revealed that these improvements were largely due to the overly aggressive post-processing used to derive hard decisions in our earlier diarization system. Therefore, in this work, we always use hard diarization decisions -- whether from the ground-truth or the real diarization outputs -- during decoding, as this approach matches the setup used during training. Note that in such a case, equation~\eqref{eq:FDDT} defining the FDDT transform reduces to a straightforward selection of one of the affine transformations corresponding to the respective STNO classes.

Incorporating soft activations into the current framework remains an open problem for future work.

\begin{table}[!tb]
    \centering
    \caption{Comparison of overall performance with ground-truth and proposed diarization system. Both ground-truth and real system results were obtained with a model trained on multiple datasets (cf. FDDT MD in~Table~\ref{tab:FDDT_full}).}
    \label{tab:real_diarization}
    \setlength{\tabcolsep}{2.5pt} 
        \small{\begin{tabular}{lcc|cc|cccc|cc}
        \toprule
          & \multicolumn{2}{c|}{AMI-sdm}  & \multicolumn{2}{c|}{NOTSOFAR-1}   & \multicolumn{4}{c|}{Libri2Mix} & \multicolumn{2}{c}{LibriCSS}  \\
         & \multicolumn{2}{c|}{test} & \multicolumn{2}{c|}{eval-small}  & \multicolumn{2}{c}{test-clean} & \multicolumn{2}{c|}{test-both} & \multicolumn{2}{c}{test}  \\
        Diarization
         &  \rotatebox[origin=c]{90}{tcp} & \rotatebox[origin=c]{90}{tcORC}  & \rotatebox[origin=c]{90}{tcp} & \rotatebox[origin=c]{90}{tcORC}  & \rotatebox[origin=c]{90}{tcp} & \rotatebox[origin=c]{90}{tcORC}  & \rotatebox[origin=c]{90}{tcp} & \rotatebox[origin=c]{90}{tcORC}  & \rotatebox[origin=c]{90}{tcp} & \rotatebox[origin=c]{90}{tcORC} \\
        \midrule
        Ground-truth
        & 17.6 & 16.7 & 19.7 & 19.1 & 6.9  & 6.9  & 15.9 & 15.9 & 8.8 & 8.8\\
        Real system
        & 25.0 & 18.2 & 33.5 & 22.6  &  8.4 & 8.3  & 20.6 & 20.5 & 11.0 & 8.9 \\
    \bottomrule
    \end{tabular}}
\end{table}

\begin{table}[!tb]
  \centering
  \caption{Diarization performance on the test set of each corpus.}
  \label{tab:DER_performance}
  \setlength{\tabcolsep}{5pt} 
  \begin{tabular}{@{}
                  l| 
                  S[table-format=2.1]|  
                  S[table-format=2.1]  
                  S[table-format=1.1] 
                  S[table-format=1.1] 
                  @{}}
  \toprule
         Dataset &
        \multicolumn{1}{c|}{DER} & 
        \multicolumn{1}{c}{Miss} & 
        \multicolumn{1}{c}{FA} &
        \multicolumn{1}{c}{Conf.} \\
  \midrule
  AMI-sdm & 17.3 & 8.7 & 3.8 & 4.8 \\
  NOTSOFAR-1 & 22.0 & 7.0 & 6.2 & 8.8 \\
  Libri2Mix (mix clean) & 4.8 & 0.3 & 4.1 & 0.4 \\
  Libri2Mix (mix both) & 10.0 & 1.2 & 8.4 & 0.4 \\
  LibriCSS & 5.5 & 3.6 & 0.4 & 1.5 \\
  \bottomrule
  \end{tabular}
\end{table}

\subsection{Extending Whisper with Joint CTC/Attention Training and Decoding}
\begin{table}[t]
    \centering
    \caption{%
Performance of the method with and without the CTC head on the NOTSFOAR-1 eval set, evaluated using tcpWER with a 5s collar. The table compares the effects of varying \( \lambda \) values in~\eqref{eq:decoding}. When \( \lambda = 1.0 \), decoding is still primarily guided by the autoregressive model, and only the top 1000 tokens are rescored by CTC. The `CTC only' column shows performance when the model uses only CTC for decoding. 
    }\label{tab:CTC_decoding}
    \small{\begin{tabular}{lccccc}
        \toprule
        & w/o CTC head & $\lambda=0$ & $\lambda=0.2$ & $\lambda=1.0$ & CTC only  \\
        \midrule
        greedy &   22.9 & 22.1 & 21.7 & 27.9 & 46.5 \\
        beam 5 &   22.2 & 21.7 & 20.9 & 32.0 & 52.8 \\
        \bottomrule
    \end{tabular}}
\end{table}

Table~\ref{tab:CTC_decoding} shows the effect of the CTC head on speaker-attributed ASR performance. First, we observe a significant improvement in tcpWER simply by adding joint training with the CTC head, even without using it during decoding (see $\lambda = 0$). This highlights the importance of guiding the training by enforcing a monotonic alignment between input frames and the output token sequence, as the Whisper decoder (or AED ASR) does not assume such alignment by default.
Furthermore, using the CTC head for joint decoding further improves the performance (see $\lambda = 0.2$), although we note that using only the CTC head for decoding is significantly worse than the Whisper decoder.
Overall, we surmise that the CTC head mitigates some of Whisper's hallucinatory tendencies despite being considerably worse as a standalone decoder.

\subsection{Does Target-Speaker Training Affect Single-Speaker Performance?}

\begin{table}[!htb]
\setlength{\tabcolsep}{2.5pt} 
    \centering
    \caption{%
        Comparison of the proposed method and original Whisper on single-speaker speech with both greedy (beam size 1) and beam-search (beam size 5) decoding. $\lambda = 0.0$ indicates no use of auxilary CTC logits during decoding despite training with auxilary CTC loss.
    }\label{tab:single_speaker_decoding}
    \small{\begin{tabular}{lcccccc}
        \toprule
        & beam & \multirow{2}{*}{$\lambda$}  & \multicolumn{2}{c}{LibriSpeech} & TED-LIUM  & VoxPopuli\\
        &  size    &                & test-clean & test-other & test & test  \\
        \midrule
        Whisper& 1& - &2.5 & 4.5 & \textbf{4.3} & 10.9 \\
        CTC head& 1 &- &3.9 & 7.3 & 8.4 & 16.1 \\
        Proposed&1 & 0.0 & 2.1 & 4.3 & 5.3 & 11.2  \\
        Proposed&1 & 0.2 & \textbf{1.9}  & 4.1 & 8.5 & 11.7  \\
        \midrule
        Whisper  & 5& -  & 2.2 & 4.3 & \textbf{4.3} & \textbf{10.0} \\
        Proposed & 5 & 0.0 & 2.1 & 4.2 & 5.0 & 11.0  \\
        Proposed & 5 & 0.2 & \textbf{1.9} & \textbf{4.0} & 7.8 & 11.2  \\
        \bottomrule
    \end{tabular}}
\end{table}
Next, we examine the performance of the proposed TS-ASR on single-speaker datasets to quantify the impact of our method on the performance of the original Whisper model on single-speaker speech. 
This experiment was conducted on LibriSpeech~\cite{librispeech}, TED-LIUM~\cite{tedlium}, and VoxPopuli~\cite{voxpopuli} evaluation sets. 
The results are presented in Table~\ref{tab:single_speaker_decoding}.

On the Librispeech test sets, we observe that the proposed model slightly outperforms the baseline Whisper model, likely due to the inclusion of the LibriSpeech training data in our fine-tuning dataset.
However, we observe performance degradation on TED-LIUM and VoxPopuli, a degradation which is especially pronounced when we incorporate the CTC head, both standalone and as part of joint decoding.
This is unsurprising, as the CTC head is trained on a much more limited set of data than the original Whisper model.
Overall, we infer that the proposed training method still maintains Whisper's abilities to handle single-talker ASR but slightly degrades its domain generalization capabilities.

\subsection{Non-Whisper Models}
\label{sec:non_whisper_models}
To verify that our method is not Whisper-specific, we took an ESPNet~\cite{watanabe2018espnet} LibriSpeech recipe\footnote{\url{https://github.com/espnet/espnet/tree/master/egs2/librispeech/asr1}} and trained attention-based encoder decoder (AED) model, namely Branchformer CTC-AED~\cite{pmlr-v162-peng22a} on LibriSpeech 960h~\cite{librispeech}. Then, we added additional FDDT parameters to the pre-trained model and fine-tuned it on the AMI-sdm dataset segmented the same way as for Whisper fine-tuning.

To show that FDDT really improves the TS-ASR performance, we performed the inference in two regimes: a) utterance-level; and b) segment-group, following the evaluation protocol from~\cite{Kanda2021MultiTalkerASR}. 

\begin{table}[!htb]
    \centering
    \caption{
        Comparison of ORC-WER of fine-tuned Branchformer inference methods on AMI-sdm.
    }
    \begin{tabular}{l|ccc}
        \toprule
        Inference Style & Baseline & with FDDT\\\hline
        utterance-level & 34.5 & 34.5   \\
        segment-group & 141.2 & 26.8 & \\\bottomrule
    \end{tabular}
    \label{tab:non_wh_models}
\end{table}

Table~\ref{tab:non_wh_models} presents ORC-WER for two models per each inference style. For baseline, we took a fine-tuned model with FDDT parameters and adjusted STNO masks such that the target speaker is always active in a given segment. We can see that in the case of utterance-level inference style, diarization information does not provide any improvement over the baseline, as the segments are usually single-speaker with occasional overlap. In the case of overlap, the STNO mask has only target-speaker and overlapped masks active, as there is no silence nor non-target speaker present in the utterance-level segments. Also, there may not be enough context for transcribing overlaps well. Furthermore, there are many segments that contain almost perfect overlap. In such a case, the model fails to properly separate the overlapped transcripts.

On the other hand, there is a significant difference between the two models in the case of segment-group inference style (114.4\% absolute). It is mainly due to the Baseline model transcribing all the speakers within the segment-group; hence, introducing many insertions. This difference proves that by adding FDDT and performing fine-tuning, we efficiently transform a single-talker model into a TS-ASR model. 
Furthermore, the difference between utterance-level and segment-group performance with FDDT para\-meters demonstrates that TS-ASR models may benefit from a longer context. Specifically, the model using FDDT parameters achieved a 7.7\,\% absolute improvement in performance on segment-group inference compared to utterance-level inference.

\section{Discussion}

Despite its promising results, the proposed approach has several limitations:

\begin{itemize}
    \item {Dependency on accurate diarization:} DiCoW’s performance is strongly influenced by the quality of the diarization system. Poor diarization, such as in high-overlap or noisy environments, can degrade the overall effectiveness of the model.
    \item {Performance on synthetic versus real data:} While the approach shows strong results on both synthetic and real-world datasets, it faces challenges adapting from synthetic benchmarks like Libri2Mix to real-world scenarios due to domain mismatches.
    \item {Scalability with increasing speaker count:} DiCoW’s computational complexity and performance are affected in scenarios with a large number of speakers, as the model processes each speaker independently.
    \item {Handling of overlapping speech:} Although the Co-Attention mechanism addresses overlapping speech to some extent, fully overlapped segments involving multiple dominant speakers remain challenging.
    \item {Limited validation on unseen conditions:} While we validated DiCoW across several datasets, further testing under diverse acoustic environments, speaker characteristics, and languages is necessary to better understand its generalization capabilities.
\end{itemize}

Addressing these limitations in future work will help refine the approach and broaden its applicability to a wider range of real-world scenarios.

\section{Conclusion}
\label{sec:conclusions}

In this study, we presented DiCoW, an approach to extend a single-speaker ASR system to perform target/multi-speaker ASR using a diarization conditioning scheme. The main contributions are:
\begin{itemize}
    \item Integration of diarization for target-speaker ASR: DiCoW directly conditions Whisper’s ASR capabilities on speaker diarization outputs, bypassing traditional speaker embeddings. This simplifies the workflow, reduces dependency on synthetic data, and improves generalization to unseen speakers.
    \item Extension of long-form ASR to multi-speaker scenarios: By building on Whisper’s long-form transcription abilities, DiCoW effectively handles overlapping speech and real-world multi-speaker recordings, enabling more reliable transcription for conversations, meetings, and other challenging audio environments.
    \item Efficient fine-tuning of pre-trained ASR models: DiCoW leverages large-scale pre-trained models like Whisper, fine-tuning them with diarization conditioning to deliver strong performance across diverse datasets. This approach minimizes training costs while achieving notable accuracy gains on real-world benchmarks such as AMI and NOTSOFAR-1.
\end{itemize}

Additionally, we extended Whisper with a ``CTC head'' to mitigate hallucinations and compared different conditioning approaches in terms of ASR performance and speed of convergence, highlighting the advantages of Frame-Level Diarization-Dependent Transformations (FDDT). We demonstrated that adapting Whisper for multi-speaker ASR does not substantially degrade its performance on single-speaker recordings, although some loss of generalization capabilities was observed.

Moreover, while most analyses were based on Whisper, we showed that the proposed method can also be successful with other models, such as AED-based ASR, demonstrating the general effectiveness of the approach.

We have successfully demonstrated that our DiCoW model achieves strong performance across various datasets when provided with ground-truth diarization.
In future work, we aim to extend the framework incorporating real speaker diarization information into training in order to reduce the performance drop that comes from training exclusively on ground-truth diarization and only using real diarization during inference.

To facilitate future comparisons and analysis, we release our code and recipes at \url{https://github.com/BUTSpeechFIT/TS-ASR-Whisper}.

\section*{Acknowledgement}
The work was supported by the Czech Ministry of Education (MoE) through OP JAK project No. ID:CZ.02.01.01/00/23\_020/0008518, Brno Ph.D. Talent Scholarship Programme, and by National Science Foundation CCRI Grant No 2120435.
Computing and data repository/services were supported by MoE through e-INFRA CZ (ID:90254) and LINDAT/CLARIAH-CZ LRI (ID:90262) respectively.

\bibliographystyle{elsarticle-num} 
\bibliography{references}

\begin{thebibliography}{10}
\expandafter\ifx\csname url\endcsname\relax
  \def\url#1{\texttt{#1}}\fi
\expandafter\ifx\csname urlprefix\endcsname\relax\def\urlprefix{URL }\fi
\expandafter\ifx\csname href\endcsname\relax
  \def\href#1#2{#2} \def\path#1{#1}\fi

\bibitem{li2022recent}
J.~Li, et~al., {Recent advances in end-to-end automatic speech recognition}, APSIPA Transactions on Signal and Information Processing 11~(1) (2022).

\bibitem{watanabe2020chime}
S.~Watanabe, M.~Mandel, J.~Barker, E.~Vincent, A.~Arora, X.~Chang, S.~Khudanpur, V.~Manohar, D.~Povey, D.~Raj, et~al., {CHiME-6 challenge: Tackling multispeaker speech recognition for unsegmented recordings}, arXiv preprint arXiv:2004.09249 (2020).

\bibitem{Yu2021M2MetTI}
F.~Yu, S.~Zhang, Y.~Fu, L.~Xie, S.~Zheng, Z.~Du, W.~Huang, P.~Guo, Z.~Yan, B.~Ma, X.~Xu, H.~Bu, {M2MeT: The ICASSP 2022 Multi-Channel Multi-Party Meeting Transcription Challenge}, in: ICASSP 2022 - 2022 IEEE International Conference on Acoustics, Speech and Signal Processing (ICASSP), 2022, pp. 6167--6171.
\newblock \href {https://doi.org/10.1109/ICASSP43922.2022.9746465} {\path{doi:10.1109/ICASSP43922.2022.9746465}}.

\bibitem{cornell2023chime}
S.~Cornell, M.~S. Wiesner, S.~Watanabe, D.~Raj, X.~Chang, P.~Garcia, Y.~Masuyam, Z.-Q. Wang, S.~Squartini, S.~Khudanpur, {The CHiME-7 DASR Challenge: Distant Meeting Transcription with Multiple Devices in Diverse Scenarios}, in: 7th International Workshop on Speech Processing in Everyday Environments (CHiME 2023), 2023, pp. 1--6.
\newblock \href {https://doi.org/10.21437/CHiME.2023-1} {\path{doi:10.21437/CHiME.2023-1}}.

\bibitem{cornell2024chime}
S.~Cornell, T.~J. Park, H.~Huang, C.~Boeddeker, X.~Chang, M.~Maciejewski, M.~S. Wiesner, P.~Garcia, S.~Watanabe, {The CHiME-8 DASR Challenge for Generalizable and Array Agnostic Distant Automatic Speech Recognition and Diarization}, in: 8th International Workshop on Speech Processing in Everyday Environments (CHiME 2024), 2024, pp. 1--6.
\newblock \href {https://doi.org/10.21437/CHiME.2024-1} {\path{doi:10.21437/CHiME.2024-1}}.

\bibitem{bhandari2024reverb}
N.~Bhandari, D.~Chen, M.~A. del R\'io~Fern\'andez, N.~Delworth, J.~D. Fox, M.~Jett\'e, Q.~McNamara, C.~Miller, O.~Novotný, J.~Profant, N.~Qin, M.~Ratajczak, J.-P. Robichaud, {Reverb: Open-Source ASR and Diarization from Rev}, arXiv preprint arXiv:2410.03930 (2024).

\bibitem{yoshioka_19_meeting_transcription}
T.~Yoshioka, I.~Abramovski, C.~Aksoylar, Z.~Chen, M.~David, D.~Dimitriadis, Y.~Gong, I.~Gurvich, X.~Huang, Y.~Huang, A.~Hurvitz, L.~Jiang, S.~Koubi, E.~Krupka, I.~Leichter, C.~Liu, P.~Parthasarathy, A.~Vinnikov, L.~Wu, X.~Xiao, W.~Xiong, H.~Wang, Z.~Wang, J.~Zhang, Y.~Zhao, T.~Zhou, {Advances in Online Audio-Visual Meeting Transcription}, in: 2019 IEEE Automatic Speech Recognition and Understanding Workshop (ASRU), 2019, pp. 276--283.
\newblock \href {https://doi.org/10.1109/ASRU46091.2019.9003827} {\path{doi:10.1109/ASRU46091.2019.9003827}}.

\bibitem{desh_21_css_multitalker}
D.~Raj, P.~Denisov, Z.~Chen, H.~Erdogan, Z.~Huang, M.~He, S.~Watanabe, J.~Du, T.~Yoshioka, Y.~Luo, N.~Kanda, J.~Li, S.~Wisdom, J.~R. Hershey, {Integration of Speech Separation, Diarization, and Recognition for Multi-Speaker Meetings: System Description, Comparison, and Analysis}, in: 2021 IEEE Spoken Language Technology Workshop (SLT), 2021, pp. 897--904.
\newblock \href {https://doi.org/10.1109/SLT48900.2021.9383556} {\path{doi:10.1109/SLT48900.2021.9383556}}.

\bibitem{Kanda2019_spkloss}
N.~Kanda, S.~Horiguchi, R.~Takashima, Y.~Fujita, K.~Nagamatsu, S.~Watanabe, Auxiliary interference speaker loss for target-speaker speech recognition, Proceedings of the Annual Conference of the International Speech Communication Association, INTERSPEECH 2019-September (2019) 236--240.
\newblock \href {https://doi.org/10.21437/Interspeech.2019-1126} {\path{doi:10.21437/Interspeech.2019-1126}}.

\bibitem{karafiat2011ivector}
M.~Karafi{\'a}t, L.~Burget, P.~Mat{\v{e}}jka, O.~Glembek, J.~{\v{C}}ernock{\`y}, {iVector-based discriminative adaptation for automatic speech recognition}, in: 2011 IEEE Workshop on Automatic Speech Recognition \& Understanding, IEEE, 2011, pp. 152--157.

\bibitem{Zili23_adapting}
Z.~Huang, D.~Raj, P.~García, S.~Khudanpur, {Adapting Self-Supervised Models to Multi-Talker Speech Recognition Using Speaker Embeddings}, in: ICASSP 2023 - 2023 IEEE International Conference on Acoustics, Speech and Signal Processing (ICASSP), 2023, pp. 1--5.
\newblock \href {https://doi.org/10.1109/ICASSP49357.2023.10097139} {\path{doi:10.1109/ICASSP49357.2023.10097139}}.

\bibitem{dehak2010front}
N.~Dehak, P.~J. Kenny, R.~Dehak, P.~Dumouchel, P.~Ouellet, {Front-end factor analysis for speaker verification}, IEEE Transactions on Audio, Speech, and Language Processing 19~(4) (2010) 788--798.

\bibitem{snyder2018x}
D.~Snyder, D.~Garcia-Romero, G.~Sell, D.~Povey, S.~Khudanpur, {X-vectors: Robust dnn embeddings for speaker recognition}, in: 2018 IEEE international conference on acoustics, speech and signal processing (ICASSP), IEEE, 2018, pp. 5329--5333.

\bibitem{wang2023wespeaker}
H.~Wang, C.~Liang, S.~Wang, Z.~Chen, B.~Zhang, X.~Xiang, Y.~Deng, Y.~Qian, {Wespeaker: A research and production oriented speaker embedding learning toolkit}, in: IEEE International Conference on Acoustics, Speech and Signal Processing (ICASSP), IEEE, 2023, pp. 1--5.

\bibitem{radford2023robust}
A.~Radford, J.~W. Kim, T.~Xu, G.~Brockman, C.~McLeavey, I.~Sutskever, {Robust speech recognition via large-scale weak supervision}, in: International conference on machine learning, PMLR, 2023, pp. 28492--28518.

\bibitem{vinnikov24_interspeech}
A.~Vinnikov, A.~Ivry, A.~Hurvitz, I.~Abramovski, S.~Koubi, I.~Gurvich, S.~Peer, X.~Xiao, B.~M. Elizalde, N.~Kanda, X.~Wang, S.~Shaer, S.~Yagev, Y.~Asher, S.~Sivasankaran, Y.~Gong, M.~Tang, H.~Wang, E.~Krupka, {NOTSOFAR-1 Challenge: New Datasets, Baseline, and Tasks for Distant Meeting Transcription}, in: Interspeech 2024, 2024, pp. 5003--5007.
\newblock \href {https://doi.org/10.21437/Interspeech.2024-1788} {\path{doi:10.21437/Interspeech.2024-1788}}.

\bibitem{Mccowan2005_ami}
I.~Mccowan, J.~Carletta, W.~Kraaij, S.~Ashby, S.~Bourban, M.~Flynn, M.~Guillemot, T.~Hain, J.~Kadlec, V.~Karaiskos, M.~Kronenthal, G.~Lathoud, M.~Lincoln, A.~Lisowska~Masson, W.~Post, D.~Reidsma, P.~Wellner, The ami meeting corpus, Int'l. Conf. on Methods and Techniques in Behavioral Research (01 2005).

\bibitem{Cosentino2020LibriMixAO}
J.~Cosentino, M.~Pariente, S.~Cornell, A.~Deleforge, E.~Vincent, \href{https://api.semanticscholar.org/CorpusID:218862876}{{LibriMix: An Open-Source Dataset for Generalizable Speech Separation}}, arXiv: Audio and Speech Processing (2020).
\newline\urlprefix\url{https://api.semanticscholar.org/CorpusID:218862876}

\bibitem{kinoshita2021integrating}
K.~Kinoshita, M.~Delcroix, N.~Tawara, {Integrating end-to-end neural and clustering-based diarization: Getting the best of both worlds}, in: Proc. ICASSP, IEEE, 2021, pp. 7198--7202.

\bibitem{bredin2023pyannote}
H.~Bredin, {pyannote. audio 2.1 speaker diarization pipeline: principle, benchmark, and recipe}, in: Proc. Interspeech 2023, 2023, pp. 1983--1987.

\bibitem{chen2020continuous}
Z.~Chen, T.~Yoshioka, L.~Lu, T.~Zhou, Z.~Meng, Y.~Luo, J.~Wu, X.~Xiao, J.~Li, {Continuous speech separation: Dataset and analysis}, in: ICASSP 2020-2020 IEEE International Conference on Acoustics, Speech and Signal Processing (ICASSP), IEEE, 2020, pp. 7284--7288.

\bibitem{shafey19_interspeech}
L.~E. Shafey, H.~Soltau, I.~Shafran, {Joint Speech Recognition and Speaker Diarization via Sequence Transduction}, in: Interspeech 2019, 2019, pp. 396--400.
\newblock \href {https://doi.org/10.21437/Interspeech.2019-1943} {\path{doi:10.21437/Interspeech.2019-1943}}.

\bibitem{kanda2022transcribe}
N.~Kanda, X.~Xiao, Y.~Gaur, X.~Wang, Z.~Meng, Z.~Chen, T.~Yoshioka, {Transcribe-to-diarize: Neural speaker diarization for unlimited number of speakers using end-to-end speaker-attributed ASR}, in: ICASSP 2022-2022 IEEE International Conference on Acoustics, Speech and Signal Processing (ICASSP), IEEE, 2022, pp. 8082--8086.

\bibitem{cornell2024one}
S.~Cornell, J.-w. Jung, S.~Watanabe, S.~Squartini, {One Model to Rule Them All? Towards End-to-End Joint Speaker Diarization and Speech Recognition}, in: ICASSP 2024-2024 IEEE International Conference on Acoustics, Speech and Signal Processing (ICASSP), IEEE, 2024, pp. 11856--11860.

\bibitem{ma2024extending}
H.~Ma, Z.~Peng, M.~Shao, J.~Li, J.~Liu, {Extending Whisper with prompt tuning to target-speaker ASR}, in: ICASSP 2024-2024 IEEE International Conference on Acoustics, Speech and Signal Processing (ICASSP), IEEE, 2024, pp. 12516--12520.

\bibitem{meng24c_interspeech}
L.~Meng, J.~Kang, Y.~Wang, Z.~Jin, X.~Wu, X.~Liu, H.~Meng, {Empowering Whisper as a Joint Multi-Talker and Target-Talker Speech Recognition System}, in: Interspeech 2024, 2024, pp. 4653--4657.
\newblock \href {https://doi.org/10.21437/Interspeech.2024-971} {\path{doi:10.21437/Interspeech.2024-971}}.

\bibitem{guo2024sq}
P.~Guo, X.~Chang, H.~Lv, S.~Watanabe, L.~Xie, {SQ-Whisper: Speaker-Querying based Whisper Model for Target-Speaker ASR}, arXiv preprint arXiv:2412.05589 (2024).

\bibitem{vaswani2017attention}
A.~Vaswani, {Attention is all you need}, Advances in Neural Information Processing Systems (2017).

\bibitem{gandhi2023distil}
S.~Gandhi, P.~von Platen, A.~M. Rush, {Distil-whisper: Robust knowledge distillation via large-scale pseudo labelling}, arXiv preprint arXiv:2311.00430 (2023).

\bibitem{watanabe2017hybrid}
S.~Watanabe, T.~Hori, S.~Kim, J.~R. Hershey, T.~Hayashi, {Hybrid CTC/attention architecture for end-to-end speech recognition}, IEEE Journal of Selected Topics in Signal Processing 11~(8) (2017) 1240--1253.

\bibitem{graves_06_ctc}
A.~Graves, S.~Fern\'{a}ndez, F.~Gomez, J.~Schmidhuber, \href{https://doi.org/10.1145/1143844.1143891}{{Connectionist temporal classification: labelling unsegmented sequence data with recurrent neural networks}}, in: Proceedings of the 23rd International Conference on Machine Learning, ICML '06, Association for Computing Machinery, New York, NY, USA, 2006, p. 369–376.
\newblock \href {https://doi.org/10.1145/1143844.1143891} {\path{doi:10.1145/1143844.1143891}}.
\newline\urlprefix\url{https://doi.org/10.1145/1143844.1143891}

\bibitem{hori_joint_2017}
T.~Hori, S.~Watanabe, J.~Hershey, \href{https://aclanthology.org/P17-1048}{Joint {CTC}/attention decoding for end-to-end speech recognition}, in: R.~Barzilay, M.-Y. Kan (Eds.), Proceedings of the 55th {Annual} {Meeting} of the {Association} for {Computational} {Linguistics} ({Volume} 1: {Long} {Papers}), Association for Computational Linguistics, Vancouver, Canada, 2017, pp. 518--529.
\newblock \href {https://doi.org/10.18653/v1/P17-1048} {\path{doi:10.18653/v1/P17-1048}}.
\newline\urlprefix\url{https://aclanthology.org/P17-1048}

\bibitem{leviathan_23_speculative}
Y.~Leviathan, M.~Kalman, Y.~Matias, \href{https://proceedings.mlr.press/v202/leviathan23a.html}{{Fast Inference from Transformers via Speculative Decoding}}, in: A.~Krause, E.~Brunskill, K.~Cho, B.~Engelhardt, S.~Sabato, J.~Scarlett (Eds.), Proceedings of the 40th International Conference on Machine Learning, Vol. 202 of Proceedings of Machine Learning Research, PMLR, 2023, pp. 19274--19286.
\newline\urlprefix\url{https://proceedings.mlr.press/v202/leviathan23a.html}

\bibitem{watanabe2018espnet}
S.~Watanabe, T.~Hori, S.~Karita, T.~Hayashi, J.~Nishitoba, Y.~Unno, N.~{Enrique Yalta Soplin}, J.~Heymann, M.~Wiesner, N.~Chen, A.~Renduchintala, T.~Ochiai, \href{http://dx.doi.org/10.21437/Interspeech.2018-1456}{{ESPnet}: End-to-end speech processing toolkit}, in: Proceedings of Interspeech, 2018, pp. 2207--2211.
\newblock \href {https://doi.org/10.21437/Interspeech.2018-1456} {\path{doi:10.21437/Interspeech.2018-1456}}.
\newline\urlprefix\url{http://dx.doi.org/10.21437/Interspeech.2018-1456}

\bibitem{polok2024targetspeakerasrwhisper}
A.~Polok, D.~Klement, M.~Wiesner, S.~Khudanpur, J.~Černocký, L.~Burget, \href{https://arxiv.org/abs/2409.09543}{{Target Speaker ASR with Whisper}} (2024).
\newblock \href {http://arxiv.org/abs/2409.09543} {\path{arXiv:2409.09543}}.
\newline\urlprefix\url{https://arxiv.org/abs/2409.09543}

\bibitem{horiguchi22_coattention}
S.~Horiguchi, Y.~Takashima, P.~García, S.~Watanabe, Y.~Kawaguchi, {Multi-Channel End-To-End Neural Diarization with Distributed Microphones}, in: ICASSP 2022 - 2022 IEEE International Conference on Acoustics, Speech and Signal Processing (ICASSP), 2022, pp. 7332--7336.
\newblock \href {https://doi.org/10.1109/ICASSP43922.2022.9746749} {\path{doi:10.1109/ICASSP43922.2022.9746749}}.

\bibitem{mosner24_interspeech}
L.~Mo\v{s}ner, R.~Serizel, L.~Burget, O.~Plchot, E.~Vincent, J.~Peng, J.~\v{C}ernock\'{y}, \href{https://www.fit.vut.cz/research/publication/13322}{Multi-channel extension of pre-trained models for speaker verification}, in: Proceedings of Interspeech 2024, Vol. 2024, International Speech Communication Association, 2024, pp. 2135--2139.
\newblock \href {https://doi.org/10.21437/Interspeech.2024-1260} {\path{doi:10.21437/Interspeech.2024-1260}}.
\newline\urlprefix\url{https://www.fit.vut.cz/research/publication/13322}

\bibitem{wolf-etal-2020-transformers}
T.~Wolf, L.~Debut, V.~Sanh, J.~Chaumond, C.~Delangue, A.~Moi, P.~Cistac, T.~Rault, R.~Louf, M.~Funtowicz, J.~Davison, S.~Shleifer, P.~von Platen, C.~Ma, Y.~Jernite, J.~Plu, C.~Xu, T.~L. Scao, S.~Gugger, M.~Drame, Q.~Lhoest, A.~M. Rush, \href{https://www.aclweb.org/anthology/2020.emnlp-demos.6}{Transformers: State-of-the-art natural language processing}, in: Proceedings of the 2020 Conference on Empirical Methods in Natural Language Processing: System Demonstrations, Association for Computational Linguistics, Online, 2020, pp. 38--45.
\newline\urlprefix\url{https://www.aclweb.org/anthology/2020.emnlp-demos.6}

\bibitem{Neumann2023MeetEval}
T.~v.~Neumann, C.~B. Boeddeker, M.~Delcroix, R.~Haeb-Umbach, {MeetEval: A Toolkit for Computation of Word Error Rates for Meeting Transcription Systems}, in: Proceedings of the 7th International Workshop on Speech Processing in Everyday Environments (CHiME 2023), 2023, pp. 27--32.
\newblock \href {https://doi.org/10.21437/CHiME.2023-6} {\path{doi:10.21437/CHiME.2023-6}}.

\bibitem{librispeech}
V.~Panayotov, G.~Chen, D.~Povey, S.~Khudanpur, {Librispeech: An ASR corpus based on public domain audio books}, in: 2015 IEEE International Conference on Acoustics, Speech and Signal Processing (ICASSP), 2015, pp. 5206--5210.
\newblock \href {https://doi.org/10.1109/ICASSP.2015.7178964} {\path{doi:10.1109/ICASSP.2015.7178964}}.

\bibitem{loshchilov2018decoupled}
I.~Loshchilov, F.~Hutter, \href{https://openreview.net/forum?id=Bkg6RiCqY7}{Decoupled weight decay regularization}, in: International Conference on Learning Representations, 2019, pp. 1--18.
\newline\urlprefix\url{https://openreview.net/forum?id=Bkg6RiCqY7}

\bibitem{han2024leveraging}
J.~Han, F.~Landini, J.~Rohdin, A.~Silnova, M.~Diez, L.~Burget, {Leveraging Self-Supervised Learning for Speaker Diarization}, arXiv preprint arXiv:2409.09408 (2024).

\bibitem{kinoshita2021advances}
K.~Kinoshita, M.~Delcroix, N.~Tawara, {Advances in integration of end-to-end neural and clustering-based diarization for real conversational speech}, in: Proc. Interspeech, 2021, pp. 3565--3569.

\bibitem{plaquet23_interspeech}
A.~Plaquet, H.~Bredin, {Powerset multi-class cross entropy loss for neural speaker diarization}, in: INTERSPEECH 2023, 2023, pp. 3222--3226.
\newblock \href {https://doi.org/10.21437/Interspeech.2023-205} {\path{doi:10.21437/Interspeech.2023-205}}.

\bibitem{chen2022wavlm}
S.~Chen, C.~Wang, Z.~Chen, Y.~Wu, S.~Liu, Z.~Chen, J.~Li, N.~Kanda, T.~Yoshioka, X.~Xiao, et~al., {Wavlm: Large-scale self-supervised pre-training for full stack speech processing}, IEEE Journal of Selected Topics in Signal Processing 16~(6) (2022) 1505--1518.

\bibitem{gulati2020conformer}
A.~Gulati, J.~Qin, C.-C. Chiu, N.~Parmar, Y.~Zhang, J.~Yu, W.~Han, S.~Wang, Z.~Zhang, Y.~Wu, et~al., {Conformer: Convolution-augmented transformer for speech recognition}, in: Proc. Interspeech, 2020, pp. 5036--5040.

\bibitem{park2019auto}
T.~J. Park, K.~J. Han, M.~Kumar, S.~Narayanan, {Auto-tuning spectral clustering for speaker diarization using normalized maximum eigengap}, IEEE Signal Processing Letters 27 (2019) 381--385.

\bibitem{Kanda2021MultiTalkerASR}
N.~Kanda, G.~Ye, Y.~Wu, Y.~Gaur, X.~Wang, Z.~Meng, Z.~Chen, T.~Yoshioka, {Large-Scale Pre-Training of End-to-End Multi-Talker ASR for Meeting Transcription with Single Distant Microphone}, in: Proceedings of Interspeech 2021, 2021, pp. 3430--3434.
\newblock \href {https://doi.org/10.21437/Interspeech.2021-102} {\path{doi:10.21437/Interspeech.2021-102}}.

\bibitem{raj23_surt2}
D.~Raj, D.~Povey, S.~Khudanpur, \href{https://doi.org/10.1109/TASLP.2023.3318398}{{SURT 2.0: Advances in Transducer-Based Multi-Talker Speech Recognition}}, IEEE/ACM Trans. Audio, Speech and Lang. Proc. 31 (2023) 3800–3813.
\newblock \href {https://doi.org/10.1109/TASLP.2023.3318398} {\path{doi:10.1109/TASLP.2023.3318398}}.
\newline\urlprefix\url{https://doi.org/10.1109/TASLP.2023.3318398}

\bibitem{zarandi_23_chubert}
M.~Fazel-Zarandi, W.-N. Hsu, {Cocktail Hubert: Generalized Self-Supervised Pre-Training for Mixture and Single-Source Speech}, in: ICASSP 2023 - 2023 IEEE International Conference on Acoustics, Speech and Signal Processing (ICASSP), 2023, pp. 1--5.
\newblock \href {https://doi.org/10.1109/ICASSP49357.2023.10096630} {\path{doi:10.1109/ICASSP49357.2023.10096630}}.

\bibitem{niu24_chime}
S.~Niu, R.~Wang, J.~Du, G.~Yang, Y.~Tu, S.~Wu, S.~Qian, H.~Wu, H.~Xu, X.~Zhang, G.~Zhong, X.~Yu, J.~Chen, M.~Wang, D.~Cai, T.~Gao, G.~Wan, F.~Ma, J.~Pan, J.~Gao, {The USTC-NERCSLIP Systems for the CHiME-8 NOTSOFAR-1 Challenge}, in: 8th International Workshop on Speech Processing in Everyday Environments (CHiME 2024), 2024, pp. 31--36.
\newblock \href {https://doi.org/10.21437/CHiME.2024-7} {\path{doi:10.21437/CHiME.2024-7}}.

\bibitem{Zhang2023}
W.~Zhang, Y.~Qian, {Weakly-Supervised Speech Pre-training: A Case Study on Target Speech Recognition}, in: Proc. INTERSPEECH 2023, 2023, pp. 3517--3521.
\newblock \href {https://doi.org/10.21437/Interspeech.2023-1280} {\path{doi:10.21437/Interspeech.2023-1280}}.

\bibitem{polok24_butjhu}
A.~Polok, D.~Klement, J.~Han, S.~Sedl\'a\v{c}ek, B.~Yusuf, M.~Maciejewski, M.~Wiesner, L.~Burget, {BUT/JHU System Description for CHiME-8 NOTSOFAR-1 Challenge}, in: 8th International Workshop on Speech Processing in Everyday Environments (CHiME 2024), 2024, pp. 18--22.
\newblock \href {https://doi.org/10.21437/CHiME.2024-4} {\path{doi:10.21437/CHiME.2024-4}}.

\bibitem{tedlium}
A.~Rousseau, P.~Del{\'e}glise, Y.~Est{\`e}ve, \href{http://www.lrec-conf.org/proceedings/lrec2012/pdf/698_Paper.pdf}{{TED}-{LIUM}: an automatic speech recognition dedicated corpus}, in: N.~Calzolari, K.~Choukri, T.~Declerck, M.~U. Do{\u{g}}an, B.~Maegaard, J.~Mariani, A.~Moreno, J.~Odijk, S.~Piperidis (Eds.), Proceedings of the Eighth International Conference on Language Resources and Evaluation ({LREC}'12), European Language Resources Association (ELRA), Istanbul, Turkey, 2012, pp. 125--129.
\newline\urlprefix\url{http://www.lrec-conf.org/proceedings/lrec2012/pdf/698_Paper.pdf}

\bibitem{voxpopuli}
C.~Wang, M.~Riviere, A.~Lee, A.~Wu, C.~Talnikar, D.~Haziza, M.~Williamson, J.~Pino, E.~Dupoux, \href{https://aclanthology.org/2021.acl-long.80}{{V}ox{P}opuli: A large-scale multilingual speech corpus for representation learning, semi-supervised learning and interpretation}, in: C.~Zong, F.~Xia, W.~Li, R.~Navigli (Eds.), Proceedings of the 59th Annual Meeting of the Association for Computational Linguistics and the 11th International Joint Conference on Natural Language Processing (Volume 1: Long Papers), Association for Computational Linguistics, Online, 2021, pp. 993--1003.
\newblock \href {https://doi.org/10.18653/v1/2021.acl-long.80} {\path{doi:10.18653/v1/2021.acl-long.80}}.
\newline\urlprefix\url{https://aclanthology.org/2021.acl-long.80}

\bibitem{pmlr-v162-peng22a}
Y.~Peng, S.~Dalmia, I.~Lane, S.~Watanabe, \href{https://proceedings.mlr.press/v162/peng22a.html}{{Branchformer: Parallel {MLP}-Attention Architectures to Capture Local and Global Context for Speech Recognition and Understanding}}, in: K.~Chaudhuri, S.~Jegelka, L.~Song, C.~Szepesvari, G.~Niu, S.~Sabato (Eds.), Proceedings of the 39th International Conference on Machine Learning, Vol. 162 of Proceedings of Machine Learning Research, PMLR, 2022, pp. 17627--17643.
\newline\urlprefix\url{https://proceedings.mlr.press/v162/peng22a.html}

\end{thebibliography}
\end{document}